\documentclass{aa}  
\usepackage[utf8]{inputenc}
\usepackage{txfonts}
\usepackage{lipsum}
\usepackage{subcaption}        
\usepackage{multirow}
\usepackage{lscape}  
\usepackage{placeins}
\usepackage{graphicx}
   \usepackage{amsthm}
\usepackage{txfonts}
\usepackage{xcolor}
\usepackage{url}
\usepackage{pifont}
\usepackage{comment}
\usepackage{amsthm}
\usepackage[colorlinks=true,
    linkcolor=blue,
    citecolor=blue,
    filecolor=magenta,      
    urlcolor=cyan]{hyperref}

\usepackage{subcaption}
\usepackage{amsmath}
\usepackage{soul}
\usepackage{multirow}
\usepackage{enumitem}
\usepackage{colortbl}

\newcommand{\lsim}{\mathrel{\rlap{\lower 3pt \hbox{$\sim$}} \raise 2.0pt \hbox{$<$}}}
\newcommand{\gsim}{\mathrel{\rlap{\lower 3pt \hbox{$\sim$}} \raise 2.0pt \hbox{$>$}}}

\begin{document}

   \title{Assessing the detectability of a realistic massive black hole binary population in LSST}

   \author{L. Bertassi \inst{1,2,3}\corrauth{l.bertassi@campus.unimib.it}   
        \and B. Sala \inst{2,3,4} \and E. Davies \inst{5} \and F. Rigamonti \inst{2,3,6} \and M. Charisi \inst{5,7} \and R. Buscicchio \inst{1,3,8,9} \and P. Severgnini\inst{2} \and M. Dotti\inst{1,2,3}
        }

   \institute{
            Universit\`a degli Studi di Milano-Bicocca, Piazza della Scienza 3, 20126 Milano, Italy
        \and
            INAF - Osservatorio Astronomico di Brera, via Brera 20, I-20121 Milano, Italy
        \and
            INFN, Sezione di Milano-Bicocca, Piazza della Scienza 3, I-20126 Milano, Italy
        \and 
            Università di Trento, Via Sommarive 14, I-38123 Trento, Italy
        \and
            Department of Physics and Astronomy, Washington State University, Pullman, WA 99163, USA
        \and
            Como Lake centre for AstroPhysics (CLAP), DiSAT, Università dell’Insubria, via Valleggio 11, 22100 Como, Italy
        \and
            Institute of Astrophysics, FORTH, GR-71110, Heraklion, Greece
      \and
	       Gran Sasso Science Institute (GSSI), I-67100 L’Aquila, Italy
        \and
            Institute for Gravitational Wave Astronomy \& School of Physics and Astronomy, University of Birmingham, Birmingham, B15 2TT, UK 
    }

   \date{Received \today}

  \abstract{
  Electromagnetic signatures of massive black hole binaries (MBHB) are expected to be found in quasar light curves. Specifically, MBHB light curves
 are expected to show periodic modulations which may originate from different physical processes. The detection of such periodicities requires algorithms suitable to model modulations with arbitrary shapes embedded in quasar red noise, in massive datasets.
  With the advent of the Vera C.
  Rubin Observatory Legacy Survey of Space and Time (LSST), such sources may finally be within reach in the next decade. In this work, we assess the performance of a recently proposed flexible detection algorithm for periodicities of arbitrary shapes with simulated LSST-like data that span its 10-year baseline. For the first time, we examine a physically motivated population of binaries and assess the effect of assumed priors on the period on the periodicity search. We study two methods to set the detection threshold: the first based on the posterior odds ratio, which assumes knowledge of the expected fraction of binaries in quasar light curves, and the second, which limits the detection of false positives. We find that the expected distribution of detected periods shows a clear peak between 5 and 10 years, depending on the specific LSST band considered. We recover a detected binary fraction ranging from $\sim 20\%$ to $\sim 3 \%$ depending on the detection threshold assumed and the band analysed, with the peak of the observed period distribution depending on the underlying binary period distribution. Therefore, the detection of a population of binaries in LSST can constrain the binary population and shed light on their evolution.}


   \maketitle
\nolinenumbers

\section{Introduction}\label{sec:introduction}
Massive black hole binaries (MBHBs) are expected to form as a consequence of the hierarchical growth of galaxies  \citep[e.g.][]{Begelman1980}. 
Depending on their masses and orbital separations, these systems are expected to emit gravitational waves (GWs) in the frequency ranges probed by ongoing Pulsar Timing Array (PTA) experiments \citep[see e.g.][]{pta}, by future space-based interferometers such as the Laser Interferometer Space Antenna \citep[LISA][]{lisa1, lisa2}, and by Moon-based GW interferometers such as the Lunar Gravitational Wave Antenna \citep[LGWA, see][]{lgwa}.

The GW detection rate from MBHBs remains highly uncertain. This is due to the wide range of physical scales involved in their evolution\footnote{From $\gsim 10$ kpc to $\lsim 10^{-3}$ pc, with the exact range depending on the massive black hole (MBH) and host galaxy properties.} towards coalescence and the complex interaction between MBHs and their environment \citep[see e.g.][and references therein]{Singh_2026, Izquierdo}. The detection of MBHBs through electromagnetic observations can constrain the MBHB evolutionary path, the expected rate of GW mergers, and shed light on the PTA stochastic GW background \citep[see][]{Antoniadis23, Agazie23, Reardon23, Xu23}. 
This task is highly non-trivial, as only MBH pairs at separations from kiloparsecs (e.g. \citealt{Scialpi2024}, see \citealt{Derosa2019} for a review) down to a few parsecs \citep[e.g.][]{Rodriguez2006} can be directly resolved with current or upcoming facilities \citep[e.g.,][]{Severgnini2026}. The same is not feasible for smaller-separation binaries, for which a plethora of indirect methods have been proposed \citep[see][for recent reviews]{Bogdanovic2021, DOrazio2023}.

Upcoming time-domain surveys, like the Legacy Survey of Space and Time (LSST) of the Vera C. Rubin Observatory \citep[see][]{Ivezic2019} and the High Latitude Time Domain Survey at the Nancy Grace Roman Space Telescope \citep[see][]{haiman23}
will provide unprecedented sky coverage, cadence, and multi-band photometry. In this context, one of the most promising signatures for the presence of a small-scale \citep[$\lesssim 10^{-2}$ pc; e.g.][]{Charisi2022} MBHB is the detection of a periodic modulation in AGN light curves.
Such modulations are expected to originate from various physical processes. First, the interaction between a small-scale binary and the surrounding gas can drive periodic accretion, resulting in periodic modulations of the observed emission \citep[see][]{MacFayden2008, Hayasaki2008}. Second, when the binary separation is small enough, relativistic effects become important: Doppler boosting beams the emission along the direction of motion, producing a lighthouse-like effect \citep[see][]{DOrazio2015, Charisi2018}. Third, if the MBHB is observed nearly edge-on, gravitational lensing can periodically magnify the emission from the minidiscs around each black hole \citep[see][]{DOrazio2018, Kelley21}. The shapes of the periodic modulations due to such processes are expected to be very different: gravitational self-lensing produces narrow, spike-like features in the light curve \citep[see][]{Hu2020} while periodic accretion generally yields asymmetric and irregular profiles \citep[see][]{Duffel2020, Zrake_2021, Westernacher_Schneider22, Cocchiararo24}. Only Doppler boosting from nearly circular binaries is expected to produce approximately sinusoidal modulations \citep[see][]{DOrazio2015}.

Detecting such modulations is complicated by the intrinsic variability of AGN \citep[see][]{Ulrich1997}. Such variability is well described as red noise \citep[see][for a recent overview]{Paolillo2025}, a stochastic process with a power spectral density decreasing with increasing frequency. 
Such intrinsic noise introduces long-term correlation in the observed light curve, mimicking periodic signals when the observational baseline spans only a few apparent cycles of the underlying modulation \citep[see][]{Vaughan16, Witt2022, Robnik24}. 

Standard periodicity identification methods, such as the Lomb-Scargle periodogram analysis \citep[see][for a review]{VanderPlas18}, have been shown to struggle when the shape deviates from being sinusoidal and in the presence of red noise \citep[see][]{Lin2026}. For this reason, more robust algorithms, leveraging, for example, Gaussian processes \citep[GPs, e.g.][]{Covino2022, rigamonti25, Foustoul2025}, matched filtering \citep[e.g.][]{Park2026} and coherence-based searches \citep[e.g.][]{Kovacevic2026} have been proposed. Such methods rely either on the assumption of dealing with a sinusoidal modulation or on knowledge of the specific shape of the modulation (e.g. the matched filtering algorithms). Recently, \citet[][henceforth referred to as Paper 1]{bertassi2026} proposed a pipeline that allows the detection of periodicities with arbitrary shapes, expanding on previous GP-based algorithms. This is particularly important as periodic signals from MBHBs are rarely expected to be sinusoidal \citep[e.g.][]{DOrazio2018, Cocchiararo24}.

The goal of this work is to determine the expected fraction of detectable MBHBs by LSST assuming a realistic underlying population of binaries through the application of the analysis proposed in Paper 1. The effect of a realistic binary population has been neglected in previous studies, which assumed broad and uninformative distributions of the binary periods \citep[e.g.][]{Vaughan16, Witt2022, Banaszak2026}. Incorporating the effects of realistic period distribution is necessary, as previous studies \citep[see, e.g.][and Paper 1]{Witt2022, Lin2026} found higher detection rates for shorter-period binaries, whereas the population is expected to be dominated by longer-period binaries.

In this work, we consider an underlying period distribution reflecting what is expected from a population of circular binaries with a GW-driven separation evolution, i.e., with a period distributed as $dN/dP \propto |dP/dt|^{-1} \propto P^{5/3}$. 
This is an approximation, since binary evolution --- except for the latter stage ---  may be driven by a more complex combination of physical processes \citep[see e.g.][]{Dotti2012}, as well as non-zero eccentricity. Nevertheless, this assumption allows us to estimate the most probable periods at which GW-driven binaries are expected to be detected. To explore the dependence of our results on the assumed population, we also considered two alternative examples where the period distribution follows a $P^{1/3}$ and a $P^{8/3}$ power law, thus exploring two scenarios with flatter or steeper period distributions than the one expected for GW-driven circular binaries. Though these distributions have no direct physical interpretation, they can be used to understand how the results are sensitive to the underlying population.

Here we consider only single-band analyses, study each of the six LSST bands independently, and report the corresponding results separately. We explore two ways to calibrate the threshold for classifying an object as a binary. The first method is based on the posterior odds ratio, while the second method limits the number of false positives. 

In Section~\ref{sec:methodology}, we outline the methodology of our work, describe the algorithm, the simulations considered, and motivate the threshold choices to detect a binary. In Section~\ref{sec:results}, we present the results of the analyses showing how the efficiency and binary detection fraction depend on the chosen thresholds, priors and bands considered. Furthermore, we identify the most likely period at which binaries will be found with LSST.  Finally, in Section~\ref{sec:conclusions} we summarise our findings and draw our conclusions.

\section{Methodology} \label{sec:methodology}
To examine the periods at which binaries will be more easily identifiable in LSST, we apply a modified version of the algorithm proposed in Paper 1 to mock data. We perform a Bayesian model comparison between two models: one including the intrinsic stochastic variability of a single-MBH (\texttt{sin} hereinafter), and another including both a periodic signal and a stochastic contribution associated with a binary scenario (\texttt{bin} hereinafter). The description of how we perform the model selection is in Section~\ref{sec:model_comp}, the likelihoods for the two models considered here are discussed in Section~\ref{sec:likelihoods}, while the assumed priors are discussed in Section~\ref{sec:priors}. Finally, the mock data generation (Davies et al. 2026, in prep) is discussed in Section~\ref{sec:data}.

\subsection{Model selection and evidence computation}\label{sec:model_comp}
The selection between the \texttt{sin} and \texttt{bin} models is fully Bayesian, based on the evidence of the two competing models: 

\begin{equation}    
Z(D|M)=\int {\rm d}\boldsymbol{\theta} L\left(D|M,\boldsymbol{\theta}\right) \pi \left(\boldsymbol{\theta}\mid M\right)  ,
\label{eq:evidence}
\end{equation}
which is a function of the data ($D$), conditioned on the chosen model ($M$, either \texttt{sin} or \texttt{bin}). The evidence is defined as $L(D|M,\boldsymbol{\theta})\pi (\boldsymbol{\theta}|M)$ integrated over the parameters $\boldsymbol{\theta}$, with $L(D|M,\boldsymbol{\theta})$ being the likelihood of the model and $ \pi (\boldsymbol{\theta}|M)$ being the prior assumed on the model parameters.

To compute the evidence for each model, we use nested sampling \citep[see][]{Skilling2006}. Specifically, we leverage the implementation of \texttt{Jaxns} \citep[see][]{albert2020}: a high-performance nested sampling package based on \texttt{JAX} \citep[see][]{Bradbury2018}, together with \texttt{tinygp} \citep[see][]{tinygp}, a package for fast and optimised GP likelihood evaluation in \texttt{JAX}. Compared to the implementation presented in Paper 1, \texttt{JAX}~-based libraries enable straightforward GPU acceleration and efficient parallelisation across multiple CPU cores. This reduces the computational time by approximately a factor of 10 on average, depending on the properties of the light curve (e.g., the number of points and photometric uncertainties\footnote{We note that, while the computational cost of a single likelihood evaluation is primarily determined by the number of data points, the number of likelihood evaluations required for convergence also depends on the light curve quality. In our analyses, larger uncertainties generally result in faster convergence and broader posteriors.}). 
We run the nested sampling with 1000 live points and stop the sampler when the estimated evidence still to be gained is less than 1\%. 

Once the evidence of each model is estimated, the most probable model is chosen based on the odds ratio:
\begin{equation}
    O \equiv \frac{p({\rm \texttt{bin}}|D)}{p({\rm \texttt{sin}}|{D})} = \frac{\pi({\rm \texttt{bin})}}{\pi({\rm \texttt{sin}})} \frac{Z({D|\rm \texttt{bin}})}{Z({D|\rm \texttt{sin}})} \sim \frac{f_{\rm{bin}}}{1-f_{\rm{bin}}} B
\end{equation}
where $f_{\rm{bin}}$ is the assumed fraction of binaries in quasars and $B\equiv Z({D|\rm \texttt{bin}})/Z({D|\rm \texttt{sin}})$ is the Bayes factor, which quantifies how strongly the data and the adopted hyperparameter priors support one model over the other. The Bayes factor is weighted by the ratio between the a priori probabilities of the two models $\pi({\texttt{bin}}|D)/\pi({\texttt{sin}}|D)$, which is often assumed to be 1. In this case, we assume it to be proportional to the true ratio between MBHBs and single MBHs. Using the odds ratio is particularly useful when dealing with rare sources. Indeed, if the expected fraction of binaries is small, having a high Bayes factor may not imply that the source is really a binary. The odds ratio instead naturally encodes this information. 

We classify a source as an MBHB candidate when $O>1$, corresponding to a Bayes factor threshold of
\begin{equation}
    B>\frac{1- f_{\rm{bin}}}{f_{\rm{bin}}} \ .
    \label{eq:B_trs_odds}
\end{equation}
As the actual fraction of binaries (i.e., bona fide periodic sources) in LSST light curves (regardless of whether any given analysis can identify them as periodic) is highly uncertain\footnote{We note that theoretical predictions on such fraction expected it to range from $f_{\rm{bin}}<10^{-3}$  to a few per cent \citep[e.g.][]{Volonteri2009, Casey_Clyde2025}.},
in the following we report results for four values for $f_{\rm bin}: 10^{-1}, \ 10^{-2}, \ 10^{-3},\ 10^{-4}$. 

Given this uncertainty, we also consider an alternative criterion independent of the assumed MBHB fraction. Following, e.g., \cite{Waddell2024}, we determine the Bayes factor threshold by setting the false positive rate to $10^{-3}$ (see Section~\ref{sec:fp_trs_res}).  Here, false positives are light curves generated without a periodic component that are incorrectly classified as periodic by the algorithm\footnote{Using a prior-informed threshold as in eq.~\ref{eq:B_trs_odds}, $f_{\rm FP}$ depends on the properties of the light curves, and therefore band-dependent, as quantified in Section~\ref{sec:odds_trs_res}}.

\subsection{Likelihoods and hyperparameter estimation}\label{sec:likelihoods}
As in Paper 1, both intrinsic variability and periodic modulations are modelled through GPs. The main difference between the two models is the covariance function of the multivariate Gaussian likelihood of the GP \citep[defined as in ][]{Rasmussen2006}. Specifically, we describe the intrinsic variability of AGN with a DRW process, the covariance function of which is \citep[see][]{Kelly2009}:  
\begin{equation}
    k_{\texttt{sin}}(t_i,t_j)=\frac{1}{2}\hat{\sigma}^2 \tau \exp \left(-\frac{|t_i-t_j|}{\tau}\right)
    \label{eq:DRW_covariance},
\end{equation}
where $\tau$ is the damping timescale of the process (i.e., the light curve autocorrelation timescale in the absence of any periodic modulation), and $\hat{\sigma}$ represents the variability amplitude, quantifying the typical magnitude of intrinsic variations. 

The covariance function for the periodic plus intrinsic variability hypothesis reads instead:
\begin{equation}
    k_{\texttt{bin}}(t_i,t_j)=A^2 \exp\left(-\frac{2\sin^2\left(\frac{\pi}{P} | t_i - t_j | \right)}{l^2}\right)+
    \frac{1}{2}\hat{\sigma}^2 \tau \exp \left(-\frac{|t_i-t_j|}{\tau}\right)
    \label{eq:periodic_kernel}
\end{equation}
where $A$ is the autocorrelation strength, $P$ is the observed modulation period, and $l$ describes how rapidly the correlation decays as the time interval between two points increases within a period, allowing the algorithm to capture periodic signals that deviate from simple sinusoids, as, for example, sawtooth-shaped or self-lensing spikes. As shown in Paper 1, such a method recovers $\sim 65 \%$ of non-sinusoidal signals (i.e. the most plausible for realistic MBHBs) in idealised LSST-like mock light curves.

\subsection{Priors}\label{sec:priors}
In this paper, we explore, for the first time, the effect of priors on the performance of the periodicity search algorithm. In Table~\ref{tab:priors}, we report the assumed prior for all the model parameters, including the shape of the prior distribution, its range and the associated model. Most priors are log-uniform over the reported range. Only for $P$ and $\tau$ do we explore different distributions, as reported in Table~\ref{tab:priors}. Specifically, we assume different power-laws for the period prior, as those are associated with faster (or slower) evolution of the underlying population. For $\tau$, we use a log-normal distribution resembling real observations \citep[see][]{MacLeod10}.
\renewcommand{\arraystretch}{1.5}
\begin{table*}[h!]
    \centering
    \caption{List of priors used in this work.}
    \renewcommand{\arraystretch}{1.5}
    \begin{tabular}{|c|c|c|c|}
        \hline
        Parameter & Prior & Range & Model\\
        \hline
        \multirow{4}{*}{$P$}
        & $\pi(P) \propto P^{-1}$
         & [3 - $10^4$] & \texttt{bin}\\
        
        & $\pi_2(P) \propto P^{5/3}$
        & [3 - $10^4$] & \texttt{bin}\\
        
        & $\pi_3(P) \propto P^{1/3}$
         &[3 - $10^4$] & \texttt{bin}\\
        
        & $\pi_4(P) \propto P^{8/3}$
        & [3 - $10^4$] & \texttt{bin}\\
        \hline
        $l$ & $\pi(l) \propto l^{-1}$
        & [$10^{-6}$ - $10^2$] & \texttt{bin}\\
        \hline
        $A$ & $\pi(A) \propto A^{-1}$
        & [$10^{-5}$ - $10^2$] & \texttt{bin}\\
        \hline
        \multirow{2}{*}{$\tau$}
        & $\pi(\tau) \propto \tau^{-1}$
        & [3 - $10^4$] & \texttt{sin, bin}\\
        
        & $\pi_5(\log_{10}\tau)
        \propto \exp[-(\log_{10}\tau-2.3)^2]$
        & [3 - $10^4$] & \texttt{sin, bin}\\
        \hline
        $\hat{\sigma}$
        & $\pi(\hat{\sigma}) \propto \hat{\sigma}^{-1}$
        & [$10^{-5}$ - $10^2$] & \texttt{sin, bin}\\
        \hline
    \end{tabular}
    \tablefoot{Summary of the priors adopted for the DRW and periodic model parameters. Different sets of prior choices (used in various analyses) are indicated with a subscript. Parameters for which different priors are considered across the analysis sets are listed in multiple rows, corresponding to the different choices of prior. The range of temporal parameters (P and $\tau$) is listed in days, while the amplitude-related parameters (A and $\hat{\sigma}$) are listed in mags. The parameter $l$ is unitless}
    \label{tab:priors}
\end{table*}
\begin{table}[h!]
    \centering
    \caption{Sets of priors used in this work}
    \begin{tabular}{|c|c|}
    \hline
        Set & Priors \\
        \hline
        1 & $\pi(P),\ \pi(l),\ \pi(A),\ \pi(\tau),\ \pi(\hat{\sigma})$ \\
        \hline
        2 & $\pi_2(P),\ \pi(l),\ \pi(A),\ \pi(\tau),\ \pi(\hat{\sigma})$ \\
        \hline
        3 & $\pi(P),\ \pi(l),\ \pi(A),\ \pi_5(\tau),\ \pi(\hat{\sigma})$ \\
        \hline
        4 & $\pi_2(P),\ \pi(l),\ \pi(A),\ \pi_5(\tau),\ \pi(\hat{\sigma})$ \\
        \hline
        5 & $\pi_3(P),\ \pi(l),\ \pi(A),\ \pi(\tau),\ \pi(\hat{\sigma})$ \\
        \hline
        6 & $\pi_4(P),\ \pi(l),\ \pi(A),\ \pi(\tau),\ \pi(\hat{\sigma})$ \\
        \hline
    \end{tabular}
    \tablefoot{Summary of the prior combinations adopted in this work. The sets differ only in the priors on the time-related quantities, $P$ and $\tau$, chosen to reflect either different physical scenarios ($P$) or observational constraints ($\tau$).}
    \label{tab:sets}
\end{table}

Unlike Paper 1, where the time axis was standardised across light curves, here priors on time-related quantities are set with upper limits much larger than 10 years (i.e., larger than the considered baseline). We note that, while the time axis of all the light curves remains unchanged, we rescale the observed magnitudes by subtracting the mean value of the light curve and dividing by its standard deviation. The upper limits of the priors related to the lengthscale $l$ differ from the implementation of Paper 1, as the sampler rarely explored regions corresponding to lengthscales larger than $10^2$.\footnote{We note that though the light curves used in this work are different from the ones in Paper 1, this choice is related to the shape of the kernel itself rather than properties of the data. As noted in Paper 1, the periodic kernel assumed here progressively suppresses higher Fourier components at increasingly large values of the length scale $l$, so that the signal converges to a sinusoid regardless of the light curve being fit. Past this point, $l$ becomes degenerate with the signal's amplitude and is no longer constrained by the data. From previous analyses, this regime is already well described at the upper limit of the assumed prior.}

In this work, we combine the priors reported in Table~\ref{tab:priors} into different sets, effectively resulting in different probabilistic models.
Prior combinations corresponding to different sets summarised in Table~\ref{tab:sets} are as follows:

\begin{enumerate}
    \item Set 1: The priors are fully uninformative and follow a log-uniform distribution over the reported ranges.
    \item Set 2: The prior on the period goes as $P^{5/3}$, following a distribution of a stationary population of circular MBHBs with constant masses and orbits tightening under the effect of GW only. The priors on the other GP parameters remain fully uninformative (i.e. log-uniform in the reported ranges).
    \item Set 3: The prior distribution on $\tau$ follows the observed distribution of damping timescales in type 1 AGN \citep[see][]{MacLeod10}, which we approximate as a truncated log-normal.
    \item Set 4: The priors on $P$ and $\tau$ are informed as in sets 2 and 3, while the other priors are as in Set 1.
\end{enumerate}
 
While a full model comparison including physically motivated evolution of MBHB eccentricity and semi-major axes with and without MBHB-environment interaction is beyond the scope of this investigation, we tested two additional sets of priors to gauge how much the results are sensitive to the distribution of $P$:
\begin{enumerate}[resume]
    \item Set 5: The prior on the period reflects an accelerated evolution of the MBHB semimajor axis at large distances, i.e. a distribution of periods that goes as $P^{1/3}$.
    \item Set 6: The prior on the period reflects a decelerated evolution of the MBHB semimajor axis at small distances, i.e. a distribution of periods that goes as $P^{8/3}$.
\end{enumerate}

To reduce the computational cost, we run the nested sampler for both ${\texttt{bin}}$ and ${\texttt{sin}}$ models assuming only uninformative priors (Set 1). The evidence associated with the remaining prior sets (see Table~\ref{tab:sets}) is approximated by re-weighting the nested sampling results following \cite{Vehtari2021}\footnote{We tested this procedure on a subsample of the light curves across the range of periods considered in this work for which we also performed the full nested sampling, verifying that the reweighted evidence ratios agree with the directly estimated ones within $1\sigma$ of the combined uncertainty reported by the nested sampler.}:
\begin{equation}
\frac{Z_{\pi_i}}{Z_{\pi_0}} \approx \frac{1}{N_s} \sum_{\tilde{\theta}} \frac{\pi_i(\tilde{\theta})}{\pi_0(\tilde{\theta})}
\label{eq:reweighted_evidences}
\end{equation}
where $N_s$ is the number of samples, and $\tilde{\theta}$ are posterior samples of the model hyperparameters. The derivation of Equation~\ref{eq:reweighted_evidences} is explained in more detail in Appendix~\ref{app:Z_reweight}.

\subsection{Mock data generation} \label{sec:data}
The mock light curves used to test the search algorithm are designed to reproduce the observing strategy expected for LSST (Davies et al., in prep.). We consider 20000 sources, where 10000 are pure-noise realisations, and 10000 show a periodic modulation. All 20000 sources present a 10-year baseline, corresponding to the nominal duration of the survey. For each source, observations are generated in the six LSST bands, with a cadence that depends on the position on the sky, which is free to vary, and can be either in the Wide Fast Deep or Deep Drilling Field mode of LSST \citep[see][]{Ivezic2019}. The observing times are obtained using the LSST Operations Simulator (\texttt{OpSim}, baseline version 4.3)\footnote{More information about OpSim can be found at this link: \url{https://www.lsst.org/scientists/simulations/opsim}}. 

Pure-noise light curves are generated assuming a DRW as the intrinsic variability model of AGNs, while in light curves that show both intrinsic variability and a periodicity, the periodic signal can be composed of contributions due to different physical processes. Specifically, we consider periodic signals arising from Doppler boosting, periodic accretion, and gravitational self-lensing, as well as combinations of these effects. Periodic modulations are generated using the physical binary parameters (e.g. mass, mass ratio, eccentricity, orbital period and redshift) as inputs, randomly drawn from broad distributions. To ensure the binary signal can be realistically observed and detected by LSST, the inputs are redrawn if they do not produce a binary signal with an average magnitude that is bright enough to be below the single-epoch limit in each band of LSST or show magnitudes lower than the saturation limit (16 mag) in all 6 bands. The periodic signals are generated following Davies et al. (2026, in preparation) using \texttt{binlite} \citep[][]{dorazio2024}, generalised to arbitrary binary mass ratio through 2D simulations from \cite{Siwek_2023}. We note that in doing so we do not reproduce a realistic population of quasars in terms of mass, redshift and luminosity. Rather, we simulate examples of light curves that are expected to be detected by LSST.

The quasar noise is modelled using the public software package \texttt{EzTao} \citep{eztao_code}, with the DRW inputs defined using correlations with the binary parameters. In this work, we use the same DRW parameters for both the noise-only and periodic+noise light curves. 
 
\begin{figure*}[h!]
    \centering
    \includegraphics[width=0.85\linewidth]{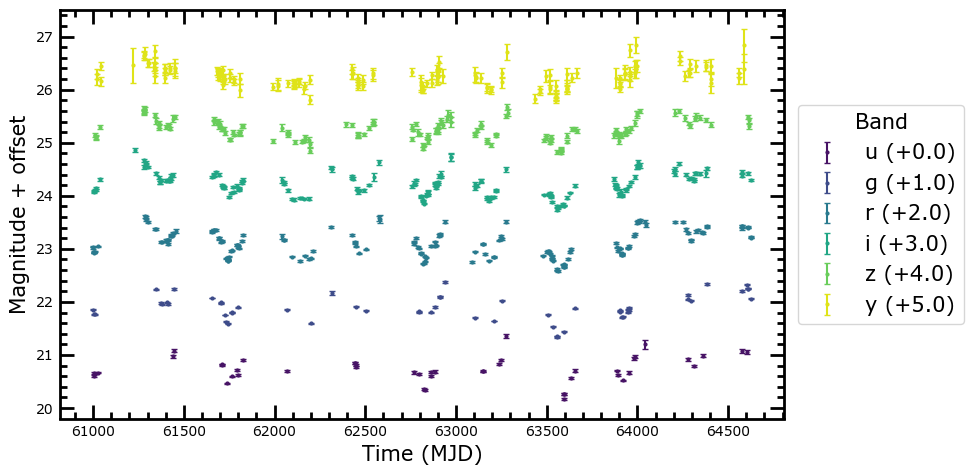}
    \caption{Example of an LSST-like light curve analysed in this work. The mean value of the time series in each band has been shifted by an offset of 1 magnitude between different bands for visualisation purposes. Different colours, from purple (u band) to yellow (y band), refer to different bands.}
    \label{fig:example_lightcurve}
\end{figure*}

In Figure~\ref{fig:example_lightcurve} we show an example of an LSST-like light curve analysed in this work. The mean value of the different bands has been shifted so that differences in cadence and SNR are more visible. 

\section{Results} \label{sec:results}

The true positive rate ($f_{\rm TP}$), defined as the fraction of correctly identified periodicities in data including a periodic signal, as well as the false positive rate, $f_{\rm FP}$, depend on the adopted periodicity detection method, the assumed value for the threshold for the Bayes factor and the (band-dependent) quality of the data. In Section~\ref{sec:odds_trs_res}, we present the results obtained by choosing a threshold based on the odds ratio ($O=1$) and deriving the threshold of the Bayes factor assuming different values of $f_{\rm bin}$, while in Section~\ref{sec:fp_trs_res}, we present the results obtained by fixing a constant $f_{\rm FP}$ for each band. Furthermore, in Section~\ref{sec:diff_pop} we study how the effect of a different underlying population affects the results.

\subsection{Odds ratio threshold} \label{sec:odds_trs_res}
\begin{figure}[h!]
    \centering
    \includegraphics[width=\linewidth]{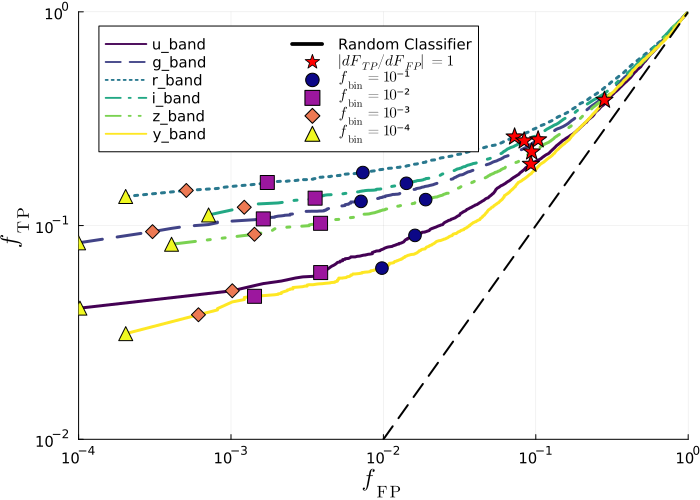}
    \caption{ROC curve considering runs with uninformative priors (set 1). Different colours, following the same colour coding as Figure~\ref{fig:example_lightcurve}, refer to different bands, which are also distinguished by linestyle: solid for the $u$ and $y$ bands,  dashed for the $g$ band, dash-dotted for the $i$ band, dash-dot-dotted for the $z$ band and dotted for the $r$ band. The black dashed line refers to a random classifier. The yellow triangles, orange diamonds, purple squares and blue circles identify the true positive fraction ($f_{\mathrm{TP}}$) and false positive fraction ($f_{\mathrm{FP}}$) at the four values of $f_{\mathrm{bin}}=10^{-4},\ 10^{-3},\ 10^{-2},\ 10^{-1}$, respectively. The red stars indicate the point where the local slope of the ROC curve is equal to unity ($|df_{TP}/df_{FP}|=1$), i.e. the point that maximises the distance from the 1-to-1 line, which is one of the thresholds used in Paper 1. We adopt logarithmic axes to have a clearer visualisation of small $f_{\rm FP}$ regions, where the points associated with the four $f_{\rm{bin}}$ considered lie.}
    \label{fig:ROC_curve}
\end{figure}

The classification power of our analysis can be assessed through the Receiver Operating Characteristic (ROC) curve, which leverages $f_{\rm{TP}}$ and $f_{\rm{FP}}$ (directly related to $f_{\rm{bin}}$ in this section). Figure~\ref{fig:ROC_curve} shows the ROC curve for the priors in set 1. For all the bands, the ROC curves lie above the 1-to-1 line over the range of thresholds considered, with $f_{\mathrm{TP}}$ consistently exceeding the corresponding false-positive fraction.

The classification performs better in the $r$ band, as it maximises the distance from the 1-to-1 line and consistently achieves the highest $f_{\rm TP}$, thanks to its dense sampling and small photometric uncertainties. The $r$ band also shows lower $f_{\rm FP}$ compared to the other bands, except for the $g$ band (and the $u$ band for $f_{\rm bin}=10^{-4}$), where the sparser sampling of the two bands tends to favour a simpler model (i.e., the DRW-only scenario) over the model with more parameters (including a periodic component) regardless of whether an actual periodicity has been included in the mock data.

In the rest of this section, we discuss the results, summarised in Table~\ref{tab:odds_ratio_TP_FP}, obtained for Sets 1 to 4. 
\begin{table*}[h!]
    \centering
    \caption{True and false positive fractions for the threshold fixed by the odds ratio}
    \begin{tabular}{|c|c|c|c|c|c|c|c|c|c|c|}
        \hline
        \multirow{2}{*}{Band} & \multirow{2}{*}{$f_{\rm{bin}}$}
          & \multicolumn{2}{c|}{Set 1}
          & \multicolumn{2}{c|}{Set 2}
          & \multicolumn{2}{c|}{Set 3}
          & \multicolumn{2}{c|}{Set 4} \\
        \cline{3-10}
        & & $f_{\mathrm{TP}}$(\%) & $f_{\mathrm{FP}}$(\%) & $f_{\mathrm{TP}}$(\%) & $f_{\mathrm{FP}}$(\%) & $f_{\mathrm{TP}}$(\%) & $f_{\mathrm{FP}}$(\%) & $f_{\mathrm{TP}}$(\%) & $f_{\mathrm{FP}}$(\%) \\
        \hline
        \multirow{4}{*}{u}
          & $10^{-1}$ & \cellcolor{blue!25}8.99 & \cellcolor{blue!25}1.61 & \cellcolor{teal!24}8.69 & \cellcolor{teal!24}2.2 & \cellcolor{purple!33}12.04 & \cellcolor{purple!33}4.4 & \cellcolor{orange!26}9.53 & \cellcolor{orange!26}2.31 \\
          & $10^{-2}$ & \cellcolor{blue!16}6.02 & \cellcolor{blue!16}0.38 & \cellcolor{teal!15}5.75 & \cellcolor{teal!15}0.26 & \cellcolor{purple!19}6.87 & \cellcolor{purple!19}0.8 & \cellcolor{orange!16}6.05 & \cellcolor{orange!16}0.40 \\
          & $10^{-3}$ & \cellcolor{blue!14}4.91 & \cellcolor{blue!14}0.10 & \cellcolor{teal!13}4.76 & \cellcolor{teal!13}0.02 & \cellcolor{purple!15}5.19 & \cellcolor{purple!15}0.26 & \cellcolor{orange!13}4.85 & \cellcolor{orange!13}0.10 \\
          & $10^{-4}$ & \cellcolor{blue!13}4.11 & \cellcolor{blue!13}0.01 & \cellcolor{teal!13}4.03 & \cellcolor{teal!13}$< 10^{-2}$ & \cellcolor{purple!13}4.28 & \cellcolor{purple!13}0.07 & \cellcolor{orange!13}4.01 & \cellcolor{orange!13}0.01 \\
        \hline
        \multirow{4}{*}{g}
          & $10^{-1}$ & \cellcolor{blue!36}12.97 & \cellcolor{blue!36}0.71 & \cellcolor{teal!37}13.56 & \cellcolor{teal!37}1.5 & \cellcolor{purple!37}13.62 & \cellcolor{purple!37}1.33 & \cellcolor{orange!36}13.36 & \cellcolor{orange!36}1.44 \\
          & $10^{-2}$ & \cellcolor{blue!29}10.74 & \cellcolor{blue!29}0.16 & \cellcolor{teal!29}10.71 & \cellcolor{teal!29}0.24 & \cellcolor{purple!30}11.02 & \cellcolor{purple!30}0.25 & \cellcolor{orange!30}10.66 & \cellcolor{orange!30}0.18 \\
          & $10^{-3}$ & \cellcolor{blue!25}9.29 & \cellcolor{blue!25}0.03 & \cellcolor{teal!25}9.11 & \cellcolor{teal!25}0.06 & \cellcolor{purple!26}9.46 & \cellcolor{purple!26}0.06 & \cellcolor{orange!24}9.08 & \cellcolor{orange!24}0.06 \\
          & $10^{-4}$ & \cellcolor{blue!23}8.3 & \cellcolor{blue!23}$< 10^{-2}$ & \cellcolor{teal!22}8.20 & \cellcolor{teal!22}0.01 & \cellcolor{purple!23}8.26 & \cellcolor{purple!23}$< 10^{-2}$ & \cellcolor{orange!22}8.16 & \cellcolor{orange!22}0.01 \\
        \hline
        \multirow{4}{*}{r}
          & $10^{-1}$ & \cellcolor{blue!48}17.70 & \cellcolor{blue!48}0.73 & \cellcolor{teal!49}18.39 & \cellcolor{teal!49}1.42 & \cellcolor{purple!48}17.83 & \cellcolor{purple!48}1.03 & \cellcolor{orange!49}17.94 & \cellcolor{orange!49}1.22 \\
          & $10^{-2}$ & \cellcolor{blue!44}15.90 & \cellcolor{blue!44}0.17 & \cellcolor{teal!44}15.87 & \cellcolor{teal!44}0.24 & \cellcolor{purple!43}15.75 & \cellcolor{purple!43}0.23 & \cellcolor{orange!43}15.63 & \cellcolor{orange!43}0.28 \\
          & $10^{-3}$ & \cellcolor{blue!40}14.53 & \cellcolor{blue!40}0.05 & \cellcolor{teal!40}14.51 & \cellcolor{teal!40}0.06 & \cellcolor{purple!40}14.53 & \cellcolor{purple!40}0.09 & \cellcolor{orange!39}14.42 & \cellcolor{orange!39}0.06 \\
          & $10^{-4}$ & \cellcolor{blue!38}13.68 & \cellcolor{blue!38}0.02 & \cellcolor{teal!38}13.63 & \cellcolor{teal!38}0.02 & \cellcolor{purple!38}13.65 & \cellcolor{purple!38}0.03 & \cellcolor{orange!37}13.50 & \cellcolor{orange!37}0.02 \\
        \hline
        \multirow{4}{*}{i}
          & $10^{-1}$ & \cellcolor{blue!43}15.76 & \cellcolor{blue!43}1.4 & \cellcolor{teal!45}16.23 & \cellcolor{teal!45}2.07 & \cellcolor{purple!45}16.25 & \cellcolor{purple!45}2.39 & \cellcolor{orange!43}15.78 & \cellcolor{orange!43}1.64 \\
          & $10^{-2}$ & \cellcolor{blue!37}13.43 & \cellcolor{blue!37}0.35 & \cellcolor{teal!37}13.34 & \cellcolor{teal!37}0.33 & \cellcolor{purple!37}13.43 & \cellcolor{purple!37}0.66 & \cellcolor{orange!36}13.11 & \cellcolor{orange!36}0.35 \\
          & $10^{-3}$ & \cellcolor{blue!33}12.13 & \cellcolor{blue!33}0.12 & \cellcolor{teal!32}11.97 & \cellcolor{teal!32}0.11 & \cellcolor{purple!33}12.1 & \cellcolor{purple!33}0.19 & \cellcolor{orange!32}11.88 & \cellcolor{orange!32}0.11 \\
          & $10^{-4}$ & \cellcolor{blue!31}11.21 & \cellcolor{blue!31}0.07 & \cellcolor{teal!31}11.14 & \cellcolor{teal!31}0.05 & \cellcolor{purple!31}11.14 & \cellcolor{purple!31}0.08 & \cellcolor{orange!30}10.96 & \cellcolor{orange!30}0.07 \\
        \hline
        \multirow{4}{*}{z}
          & $10^{-1}$ & \cellcolor{blue!36}13.22 & \cellcolor{blue!36}1.89 & \cellcolor{teal!36}13.31 & \cellcolor{teal!36}2.43 & \cellcolor{purple!38}13.91 & \cellcolor{purple!38}3.35 & \cellcolor{orange!35}12.77 & \cellcolor{orange!35}1.98 \\
          & $10^{-2}$ & \cellcolor{blue!28}10.25 & \cellcolor{blue!28}0.39 & \cellcolor{teal!28}10.17 & \cellcolor{teal!28}0.29 & \cellcolor{purple!29}10.43 & \cellcolor{purple!29}0.89 & \cellcolor{orange!27}10.01 & \cellcolor{orange!27}0.36 \\
          & $10^{-3}$ & \cellcolor{blue!25}9.06 & \cellcolor{blue!25}0.10 & \cellcolor{teal!25}8.91 & \cellcolor{teal!25}0.07 & \cellcolor{purple!25}8.98 & \cellcolor{purple!25}0.22 & \cellcolor{orange!24}8.80 & \cellcolor{orange!24}0.11 \\
          & $10^{-4}$ & \cellcolor{blue!22}8.18 & \cellcolor{blue!22}0.04 & \cellcolor{teal!21}8.09 & \cellcolor{teal!21}0.03 & \cellcolor{purple!21}8.05 & \cellcolor{purple!21}0.07 & \cellcolor{orange!20}7.90 & \cellcolor{orange!20}0.04 \\
        \hline
        \multirow{4}{*}{y}
          & $10^{-1}$ & \cellcolor{blue!17}6.32 & \cellcolor{blue!17}0.98 & \cellcolor{teal!20}7.41 & \cellcolor{teal!20}2.05 & \cellcolor{purple!20}7.35 & \cellcolor{purple!20}2.43 & \cellcolor{orange!18}6.56 & \cellcolor{orange!18}1.67 \\
          & $10^{-2}$ & \cellcolor{blue!12}4.67 & \cellcolor{blue!12}0.14 & \cellcolor{teal!12}4.63 & \cellcolor{teal!12}0.19 & \cellcolor{purple!13}4.74 & \cellcolor{purple!13}0.42 & \cellcolor{orange!12}4.46 & \cellcolor{orange!12}0.17 \\
          & $10^{-3}$ & \cellcolor{blue!10}3.79 & \cellcolor{blue!10}0.05 & \cellcolor{teal!10}3.62 & \cellcolor{teal!10}0.05 & \cellcolor{purple!10}3.69 & \cellcolor{purple!10}0.12 & \cellcolor{orange!9}3.54 & \cellcolor{orange!9}0.08 \\
          & $10^{-4}$ & \cellcolor{blue!8}3.12 & \cellcolor{blue!8}0.02 & \cellcolor{teal!8}3.04 & \cellcolor{teal!8}0.01 & \cellcolor{purple!8}3.09 & \cellcolor{purple!8}0.06 & \cellcolor{orange!8}2.93 & \cellcolor{orange!8}0.02 \\
        \hline
    \end{tabular}
    \tablefoot{True positive and false positive rates obtained by fixing the Bayes factor threshold through the Odds ratio, for different bands and prior sets (distinguished by colour: blue, teal, purple, and orange for Sets 1–4). Darker cells indicate higher $f_{\mathrm{TP}}$ and lighter cells indicate lower $f_{\mathrm{TP}}$.}
    \label{tab:odds_ratio_TP_FP}
\end{table*}
Specifically, Table~\ref{tab:odds_ratio_TP_FP} reports $f_{\mathrm{TP}}$ and $f_{\mathrm{FP}}$ for different bands (first column) in different rows ($u$, $g$, $r$, $i$, $z$, $y$ from top to bottom) and for the different prior sets in different columns. Each prior set is reported with a different colour: blue, teal, purple and orange for Sets 1, 2, 3, 4 respectively. \footnote{These are the prior sets with a direct physical interpretation, i.e a population of circular MBHB driven by GW emission (Set 2), a distribution of $\tau$ from real observations (Set 3) or both (Set 4). Sets 5 and 6 are discussed later in Section~\ref{sec:diff_pop}} For each pair of prior set and band, we report $f_{\mathrm{TP}}$ and $f_{\mathrm{FP}}$ for different $f_{\rm{bin}}$. Specifically, we report the results for $f_{\rm{bin}}=10^{-1}, \ 10^{-2}, \ 10^{-3}, \ 10^{-4}$ on different rows. We use different shading to highlight how $f_{\rm{TP}}$ changes across prior sets and bands, moving from darker colours for higher $f_{\rm{TP}}$ to lighter colours for lower $f_{\rm{TP}}$.

Looking at Table~\ref{tab:odds_ratio_TP_FP} as a whole, we notice that, independently of the assumed prior set, $f_\mathrm{TP}$ ranges between $\sim 3\%$ and $\sim 20\%$, with the $r$ band being the band with the higher $f_\mathrm{TP}$, while $f_{\rm{FP}}$ is always smaller than $4.5\%$. The values of $f_{\mathrm{TP}}$ and $f_{\mathrm{FP}}$, for Set 1 only, are highlighted as circles in the ROC curve shown in Figure~\ref{fig:ROC_curve}. Such values of $f_{\rm{TP}}$ are smaller when compared with the LSST light curves studied in Paper 1, which showed values of $f_{\rm{TP}}$ ranging between 50\% and 70\% depending on the shape of the signal. This is due to several factors:

\begin{itemize}
    \item In Paper 1, idealised LSST-like light curves, with higher cadence and smaller photometric errors, were used.
    \item In Paper 1, the threshold for detectability did not consider the underlying population of binaries and the rarity of such sources among quasars. Rather, the threshold was set either by fixing $f_{\rm{FP}}$ or by finding the Bayes factor of the most favourable point in the ROC curves, e.g. the combination of $f_{\mathrm{TP}}$ and $f_{\mathrm{FP}}$ that maximises the distance from the 1-to-1 curve (shown in Figure~\ref{fig:ROC_curve} as red stars). Setting the threshold through the ROC curve would result in a higher $f_{\rm{TP}}$ but with an increased $f_{\rm{FP}}$. 
    \item As discussed in Section~\ref{sec:data}, the light curves considered in this work present a much more complex phenomenology when compared with the sinusoids and sawtooth analysed in Paper 1.
    \item Observational uncertainties are band-dependent and, on average across bands, are larger than the ones associated with light curves in Paper 1, resulting in an overall lower Signal-to-Noise Ratio (SNR)\footnote{We define the SNR as the average of the ratio between the flux and its uncertainty} when compared to Paper 1. This effect is the primary driver of the lower $f_{\rm{TP}}$.
\end{itemize}

Focusing on a given prior set, e.g. Set 1 for which the ROC curves are shown in Figure~\ref{fig:ROC_curve}, it can be seen that $f_{\mathrm{TP}}$ decreases with decreasing values of $f_{\rm{bin}}$ (moving from darker to lighter rows in terms of colouring). In contrast, the ratio $f_{\mathrm{TP}}/f_{\mathrm{FP}}$ increases as $f_{\rm{bin}}$ decreases. This is because, as $f_{\rm{bin}}$ decreases, the Bayes factor threshold increases, reducing the number of pure-noise light curves with a Bayes factor above the threshold. At the same time, some periodic light curves survive the highest $B$ threshold. Such sources would be considered the most promising candidates once the algorithm is applied to real data. We further note that this trade-off, i.e., a lower $f_{\rm{bin}}$ allowing for a higher classification power, is independent of the assumed set of priors.

For fixed $f_{\rm{bin}}$ and independently of the assumed prior set, $f_{\rm{TP}}$ has a strong dependence on the considered band (this effect can be observed also in Figure~\ref{fig:ROC_curve}). Specifically, the $r$ and $i$ bands, being better sampled, show the highest $f_{\mathrm{TP}}$, while the less sampled bands (i.e. the $u$ band) or the bands with low SNR (e.g. the $y$ band) show the lowest $f_{\mathrm{TP}}$. Such behaviour is expected: if the light curve has too few points, the presence of long-term correlations can be more easily explained by a red-noise realisation since there is not enough information on the possibly deterministic behaviour of the light curve over short timescales. In a similar way, a low SNR does not guarantee a high enough precision on individual points to discriminate between a periodic and a noise realisation. Since the \texttt{sin} model shows a lower number of parameters, it is more easily favoured by the Bayesian model comparison in these two conditions. In the following, we report the results for the $r$ and $y$ bands, i.e. the bands with the highest and lowest $f_{\mathrm{TP}}$, respectively.

\begin{figure}[h!] 
    \centering
    \begin{subfigure}[t]{\linewidth}
        \centering
        \includegraphics[width=\linewidth]{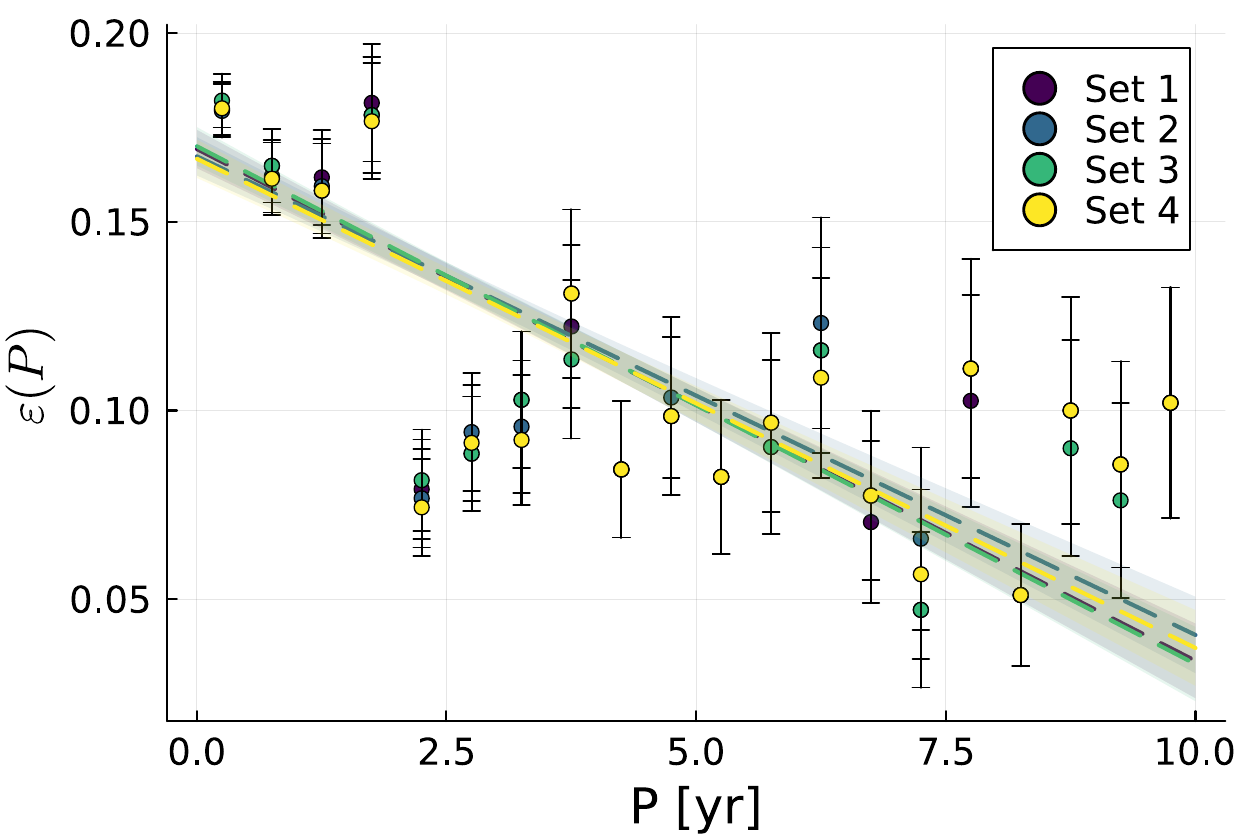}
        \caption{$r$ band}
    \end{subfigure}
    \vspace{1em}
    \begin{subfigure}[t]{\linewidth}
        \centering
        \includegraphics[width=\linewidth]{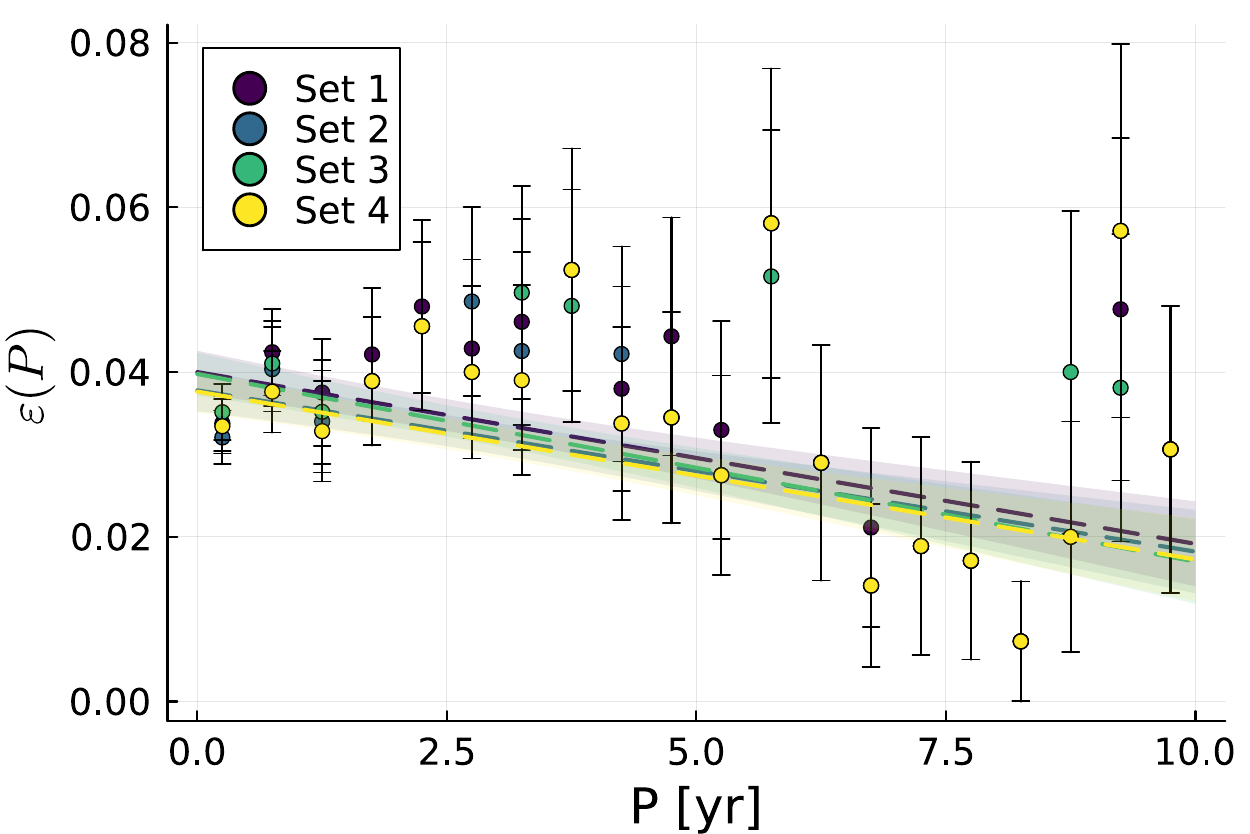}
        \caption{$y$ band}
    \end{subfigure}
    \caption{Detection efficiency as a function of the injected period, obtained by fixing the Bayes factor threshold through the odds ratio, for the $r$ and $y$ bands (in panels a and b, respectively). The purple line refers to Set 1, the blue line refers to Set 2, $P$, the green line refers to Set 3, and finally the yellow line refers to Set 4.}
    \label{fig:efficiency_all_odds_1_1000}
\end{figure}

The distribution of detected binaries as a function of the period depends on the efficiency ($\epsilon$) of detecting a specific period and on the underlying population (the effect of which is discussed in Section~\ref{sec:diff_pop}). 
In Figure~\ref{fig:efficiency_all_odds_1_1000} we show $\epsilon$ for prior sets 1 to 4, assuming an $f_{\rm bin}=10^{-3}$ and focusing on the $r$ (best) and $y$ (worst) bands. 

\begin{figure}[h!]
    \centering
    \begin{subfigure}[t]{\linewidth}
        \centering
        \includegraphics[width=\linewidth]{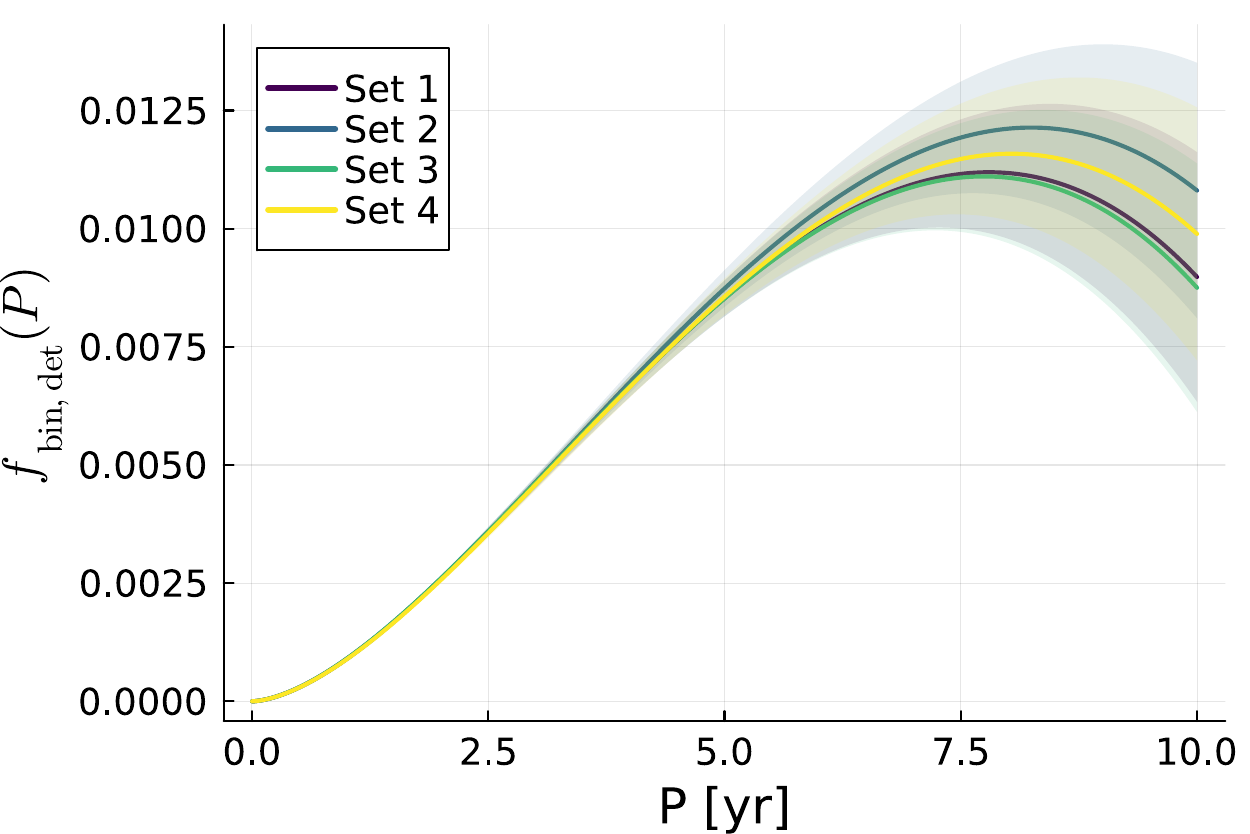}
        \caption{$r$ band}
    \end{subfigure}
    \vspace{1 em}
    \begin{subfigure}[t]{\linewidth}
        \centering
        \includegraphics[width=\linewidth]{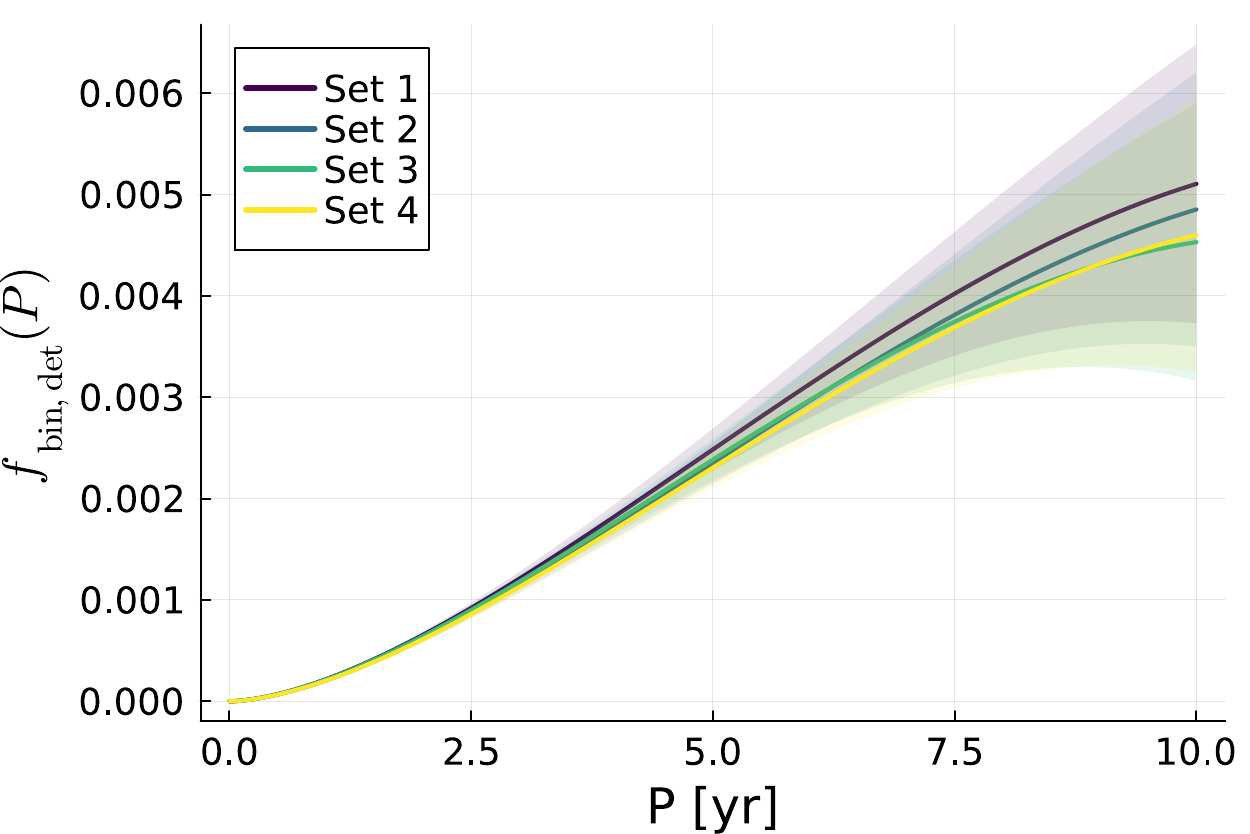}
        \caption{$y$ band}
    \end{subfigure}
    \caption{Detected fraction of binaries $f_{\mathrm{bin,det}}$ as a function of period, obtained by fixing the Bayes factor threshold through the odds ratio, for the $r$ and $y$ bands (in panels a and b, respectively). The colour scheme is the same as in Figure~\ref{fig:efficiency_all_odds_1_1000}}
    \label{fig:detection_all_odds_1_1000}
\end{figure}

To calculate $\epsilon$, we first divide the injected period distribution into bins of 0.5 yr. Then, for each bin, we calculate the ratio between the number of correctly identified binaries and the total number of binaries. The discrete fractions obtained for each period bin are linearly interpolated.\footnote{We note that such linear interpolation is a first approximation to capture the overall trend rather than finer features, which may either be due to statistical fluctuations or both physical and instrumental effects. We postpone the study of such finer structures to future work focused on the dependence of the algorithm efficiency on LSST systematics and on the physical origin of the modulation.} As can be seen in both panels of Figure~\ref{fig:efficiency_all_odds_1_1000}, the efficiency decreases as the period of the injected signal increases. This agrees with what is reported in Paper 1 and other studies \citep[e.g.][]{Witt2022, Lin2026}: the number of observed cycles
strongly affects the algorithm's ability to correctly distinguish between pure noise and periodic modulations. We also note that, from Figure~\ref{fig:efficiency_all_odds_1_1000}, the lowest $\epsilon$ in the $r$ band is higher than the highest $\epsilon$ in the $y$ band, further showing how different properties of the bands impact our ability to identify periodicities. 
Multiplying the retrieved efficiencies by the realistic GW-driven population of binaries, we retrieve the fraction of detected binaries,  $f_{\mathrm{bin,det}}$, as a function of the MBHB period $P$. We show the results for the $r$ and $y$ bands in Figure~\ref{fig:detection_all_odds_1_1000}, where we show the product of the efficiency of our algorithm and the period distribution. The distribution of binary periods is normalised so that it integrates to 1 over the considered period range, and consequently it does not depend on the total number of binaries. 

Depending on the combined effect of the efficiency and assumed period distribution, we see different behaviours in $f_{\rm{bin,det}}$. Specifically, if the efficiency drops rapidly (like in r-band), we observe a clear peak in the detected fraction. This can be easily observed in panel (a) of Figure~\ref{fig:detection_all_odds_1_1000}, where the peak is localised between 7.5 and 10 years for the $r$ band. When the efficiency decreases more slowly (e.g., in the $y$ band), i.e. the slope of the fitted line flattens due to the limited constraining power of the data, the $f_{\rm{bin,det}}$ tends to follow the underlying population, as can be seen in panel (b), where the detected fraction for the $y$ band is shown. Results similar to the $ r$ band are obtained for the $u$, $g$ and $i$ bands. Specifically, in the $g$ and $u$ bands the peak is found between 5 and 7.5 years, while in the $i$ band the peak is found between 7.5 and 10 years. The $z$ band shows results similar to the $y$ band without a clear peak in $f_{\rm{bin}}$ following the underlying period distribution. We note that the position of the peak can also depend on the minimum number of cycles required to be observed within the baseline. The effect of such a requirement will be studied in future work.

\subsection{False positive threshold} \label{sec:fp_trs_res}

Motivated by our limited knowledge of the expected binary fraction, we also consider an alternative criterion in which we fix the false-positive fraction to $f_{\rm FP}=10^{-3}$ and use it to determine a band-dependent threshold for $B$.

For each band, we consider a set of 10000 mock noise-only light curves and fit them considering the different sets of priors 1-4  described in Table~\ref{tab:sets}. This allows us to determine the distribution of Bayes factors for our mock population of single AGN and to determine the $f_{\rm{FP}}$-based $B$ thresholds, which we report in Table~\ref{tab:fractions}. Using these thresholds, we then analyse 10000 mock periodic light curves (per band) and obtain
the $f_{\mathrm{TP}}$ -- for each of the 6 bands for all four priors used in this work -- reported in Table~\ref{tab:fractions}. We note that, similar to the fractions listed in Table~\ref{tab:odds_ratio_TP_FP}, the $f_{\mathrm{TP}}$ are smaller than those reported in Paper 1 for the same reasons discussed in Section~\ref{sec:odds_trs_res}.

\begin{table*}[h!]
    \centering
    \caption{Thresholds and true positive fraction}
    \begin{tabular}{|c|c|c|c|c|c|c|c|c|}
        \hline
        \multirow{2}{*}{Band} 
          & \multicolumn{2}{c|}{Set 1} 
          & \multicolumn{2}{c|}{Set 2} 
          & \multicolumn{2}{c|}{Set 3} 
          & \multicolumn{2}{c|}{Set 4} \\
        \cline{2-9}
        & $\log_{10}B_{\rm{trs}}$ & TP(\%) & $\log_{10}B_{\rm{trs}}$ & TP(\%) & $\log_{10}B_{\rm{trs}}$ & TP(\%) & $\log_{10}B_{\rm{trs}}$ & TP(\%) \\
        \hline
        u & \cellcolor{blue!13} 2.89 & \cellcolor{blue!13} 5.01 & \cellcolor{teal!13} 2.62 & \cellcolor{teal!13} 4.98 & \cellcolor{purple!11} 3.69 &\cellcolor{purple!11} 4.51 &\cellcolor{orange!13} 2.82 &\cellcolor{orange!13} 4.98 \\
        \hline
        g & \cellcolor{blue!28} 2.11 & \cellcolor{blue!28} 10.5 &\cellcolor{teal!28} 2.36 &\cellcolor{teal!28} 10.1 &\cellcolor{purple!28} 2.63 &\cellcolor{purple!28} 9.96 &\cellcolor{orange!28} 2.37 &\cellcolor{orange!28} 9.98 \\
        \hline
        r & \cellcolor{blue!42} 2.43 & \cellcolor{blue!42} 15.3 &\cellcolor{teal!40} 2.75 &\cellcolor{teal!40} 14.8 &\cellcolor{purple!40} 2.88 &\cellcolor{purple!40} 14.7 &\cellcolor{orange!41} 2.51 &\cellcolor{orange!41} 14.9 \\
        \hline
        i & \cellcolor{blue!33} 3.11 & \cellcolor{blue!33} 12.01 &\cellcolor{teal!33} 3.07 & \cellcolor{teal!33} 11.94 &\cellcolor{purple!32} 3.55 &\cellcolor{purple!32} 11.61 &\cellcolor{orange!32} 3.11 &\cellcolor{orange!32} 11.77 \\
        \hline
        z & \cellcolor{blue!25} 2.98 & \cellcolor{blue!25} 9.1 &\cellcolor{teal!25} 2.79 &\cellcolor{teal!25} 9.1 &\cellcolor{purple!23} 3.58 &\cellcolor{purple!23} 8.37 &\cellcolor{orange!24} 3.19 & \cellcolor{orange!24}8.59 \\
        \hline
        y & \cellcolor{blue!12} 2.28 & \cellcolor{blue!12} 4.39 &\cellcolor{teal!12} 2.27 &\cellcolor{teal!12} 4.45 &\cellcolor{purple!10} 3.16 &\cellcolor{purple!10} 3.54 &\cellcolor{orange!11} 2.75 & \cellcolor{orange!11} 3.73 \\
        \hline
    \end{tabular}
    \tablefoot{Bayes factor threshold and true positive fraction obtained for a fixed false positive rate at $10^{-3}$, for different bands and prior sets, distinguished by colour: blue, teal, purple, and orange for Sets 1–4. Darker cells indicate higher $f_{\mathrm{TP}}$ while lighter cells indicate lower $f_{\mathrm{TP}}$.}
    \label{tab:fractions}
\end{table*}

\begin{figure}[h!]
    \centering
    \begin{subfigure}[t]{\linewidth}
        \centering
        \includegraphics[width=\linewidth]{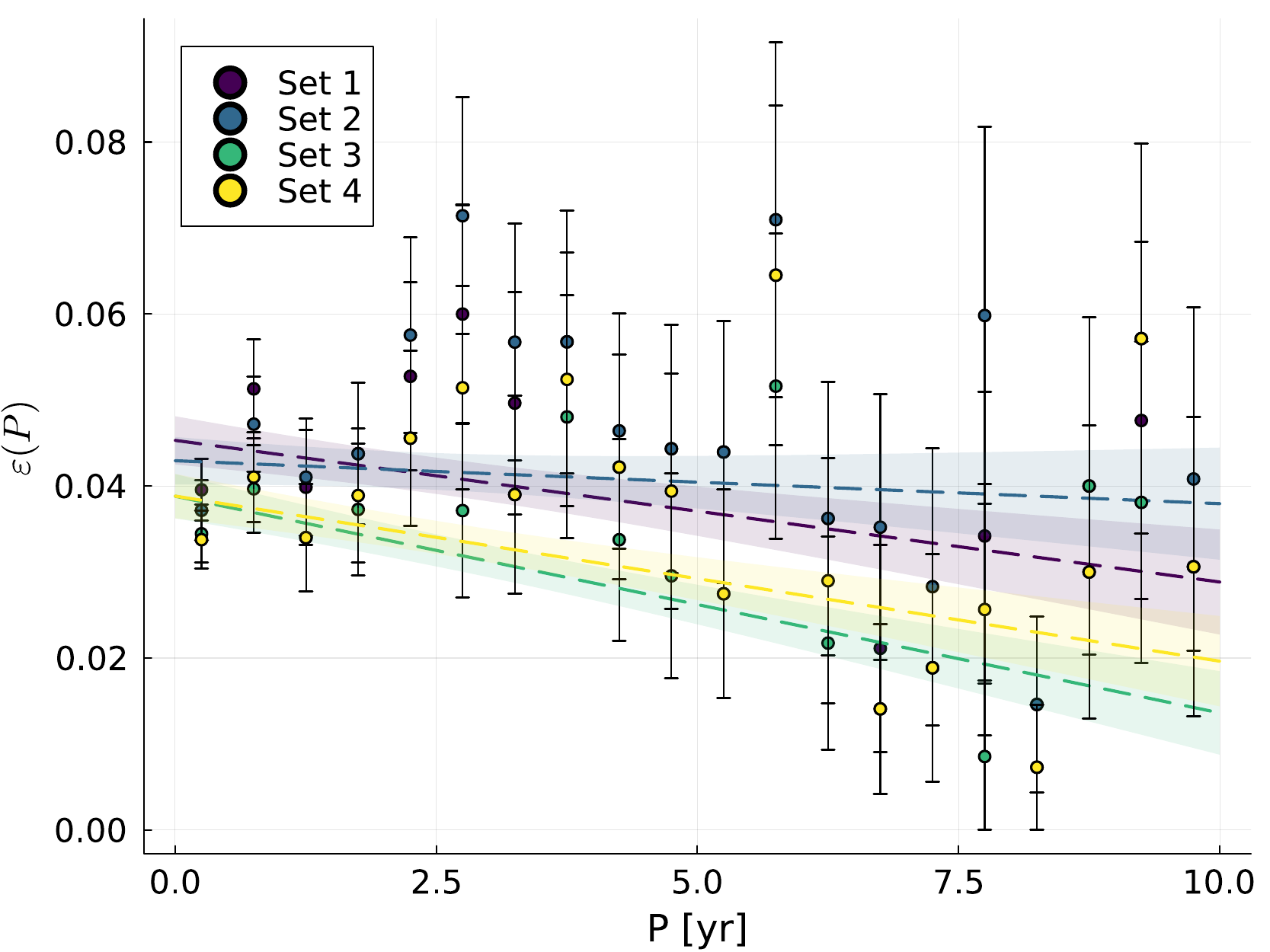}
        \caption{$\epsilon(P)$}
    \end{subfigure}
    \vspace{1 em}
    \begin{subfigure}[t]{\linewidth}
        \centering
        \includegraphics[width=\linewidth]{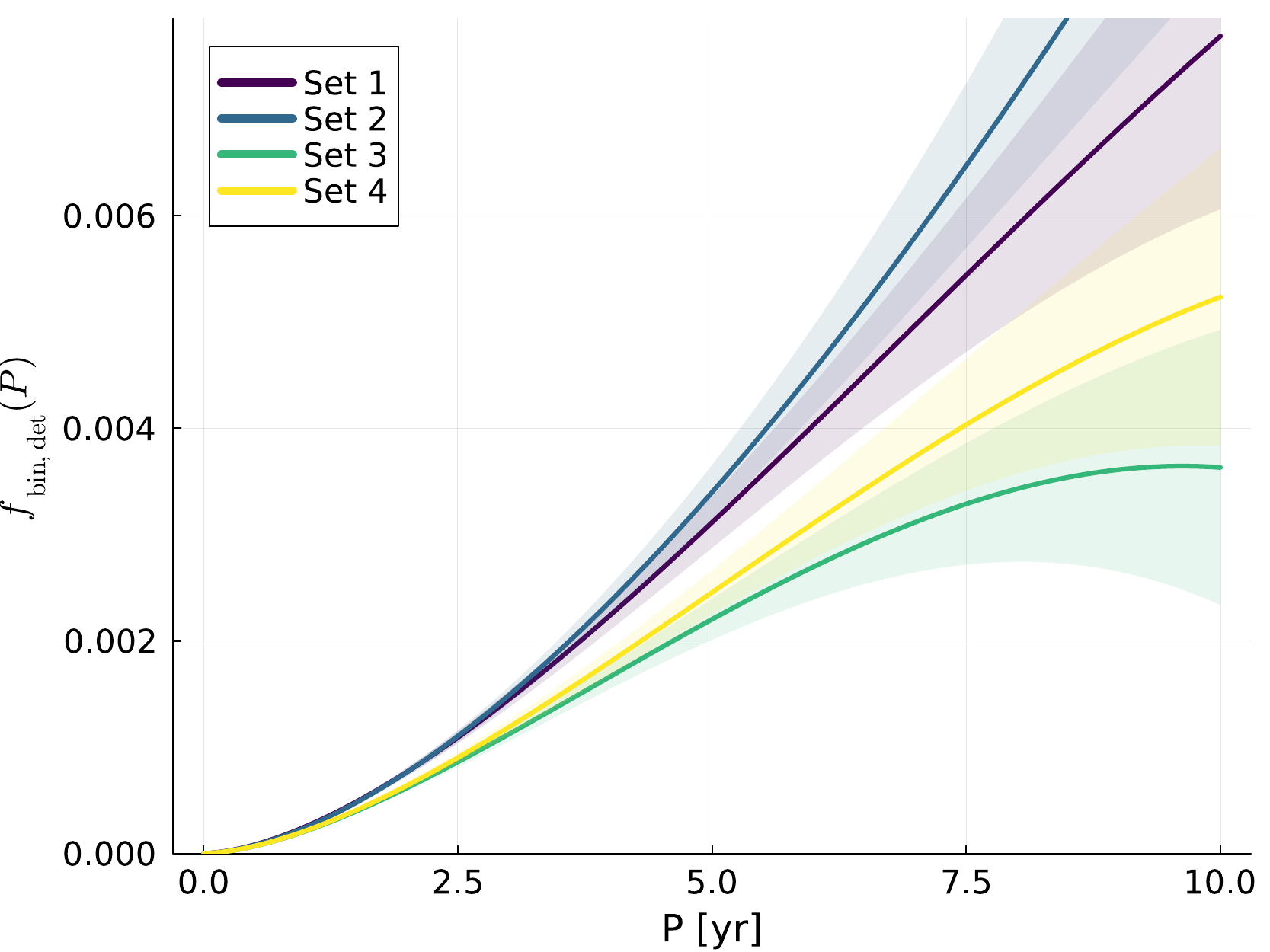}
        \caption{$f_{\rm{bin,det}}$ band}
    \end{subfigure}
    \caption{Efficiency and detected binary fraction as a function of period for the $y$ band shown in panels (a) and (b), respectively, found by fixing the threshold by limiting the number of false positives. The purple line refers to Set 1, the blue line refers to the analysis performed considering Set 2, the green line refers to analysis performed assuming Set 3 and finally the yellow line refers to analysis performed assuming Set 4.}
    \label{fig:efficiency_all}
\end{figure}

We compute $\epsilon$ as discussed in Section~\ref{sec:odds_trs_res}. The results are very similar to the ones presented in Figure~\ref{fig:efficiency_all_odds_1_1000}. Specifically, the efficiencies decrease with increasing period, with a steeper decrease in the $r$ band and a lower efficiency for the $y$ band. The main difference between the results in this section and those in Section~\ref{sec:odds_trs_res} is in the $y$ band, where the efficiency curves obtained under different priors are visibly separated, having different slopes, as shown in panel (a) of Figure~\ref{fig:efficiency_all}. This increased sensitivity to the prior choice can be explained by the low SNR of this band: with less informative data, the results are more strongly influenced by the prior. Specifically, the Bayes factor threshold is set from the distribution of the Bayes factors of the false positives, which, at low SNR, may itself be biased by a prior-dominated scenario. 

Similarly to Section~\ref{sec:odds_trs_res}, we compute $f_{\mathrm{bin,det}}$ using the threshold found by fixing $f_{\mathrm{FP}}$. As with the odds-ratio threshold, $f_{\mathrm{bin,det}}$ presents a clear peak when the decrease in the efficiency of the algorithm is faster than the increase in the number of sources as a function of the period.

The main difference with the fractions of Section~\ref{sec:odds_trs_res} is again in the y-band, for which $f_{\rm{bin,det}}$ is shown in panel (b) of Figure~\ref{fig:efficiency_all}. Similar to panel (a) of Figure~\ref{fig:efficiency_all}, the prior dominance on the results is clearly visible in the different behaviour of $f_{\mathrm{bin,det}}$. Assuming Set 1 or Set 2, $f_{\mathrm{bin,det}}$ follows the underlying population. However, when Sets 3 and 4 are considered, i.e., when also the prior of $\tau$ is changed, the detected fraction tends to flatten at high periods. Specifically, moving from a flat prior on $\tau$ to a log-Gaussian prior centred at high values favours red-noise realisations with long-term variations on timescales comparable to the injected period in the upper range studied. In this regime, a pure-noise (DRW-only) realisation can already explain the observed modulation without requiring a periodic component, reducing the number of light curves for which the Bayes factor exceeds the detection threshold.

\subsection{Different underlying populations}\label{sec:diff_pop}
As mentioned in Section~\ref{sec:introduction}, a population of binaries with a period distribution proportional to $P^{5/3}$ assumes circular binaries and a non-evolving chirp mass. 
Furthermore, we assume that GW emission is the dominant process driving binary evolution, while other physical processes may also play a role.
This is hinted at by recent PTA measurements of the PTA background, in which models that include additional mechanisms for binary evolution provide a better fit than GW-only driven evolution \citep[see, e.g.][]{Agazie23, EPTA2024}. For this reason, we study how $f_{\mathrm{bin,det}}$ changes when the underlying population is varied, in the specific case of the $r$ band.

\begin{figure}[h!]
    \centering
    \includegraphics[width=\linewidth]{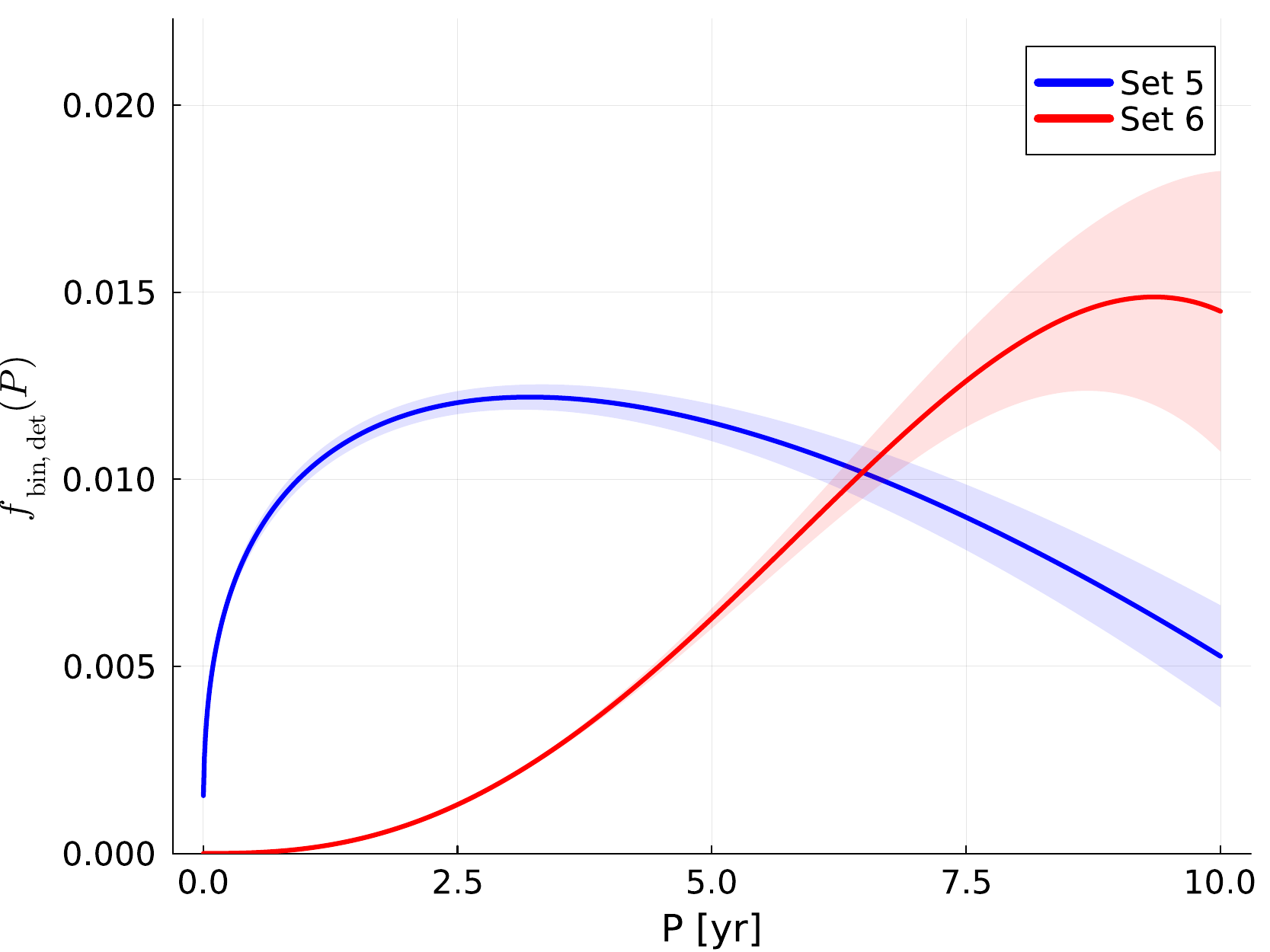}
    \caption{Detected fraction of binaries in the $r$ band assuming Set 5 (blue line) and Set 6 (red line). These fractions are found assuming the Bayes factor threshold that limits the false positive fraction to $f_{FP}=10^{-3}$. } 
    \label{fig:detect_fraction_r_8_3_1_3}
\end{figure}

In Figure~\ref{fig:detect_fraction_r_8_3_1_3}, we show the detected fraction for period distributions going as $P^{1/3}$ (blue line) and $P^{8/3}$ (red line), for the $r$ band. The overall behaviour is similar to the $P^{5/3}$ case: the combination of decreasing efficiency and an increasing number of sources with period results in a clear peak in the detected fraction. The main difference between the three cases lies in the period at which this peak occurs: the peak shifts towards higher periods as the population distribution increases more rapidly with period. This is expected, since the algorithm's efficiency is unchanged across cases, while a steeper exponent increases the relative number of binaries at longer periods. 

The shift in the most probable detectable period may have important implications, as PTA results suggest that a population of eccentric binaries is needed to explain the observed background \citep[see][]{Chen2017}. If this is the case, binary separation shrinkage will proceed faster at large separation and slow down during the GW-driven circularisation, making the underlying period population flatter. As a consequence, it will be more likely to detect a binary with a short period, which shows more than one cycle in LSST. The shape of the detected fraction can therefore be used to constrain how binaries are distributed across the range of periods (separations) considered. As soon as observations of confirmed binary sources become available, comparing the resulting distribution of detected binaries to the one predicted for the $P^{5/3}$ case could help unveil the dynamical processes contributing to the MBHB shrinking, and it would be a completely independent probe of the GW background. We note that this specific detected-fraction distribution is specific to LSST and the adopted detection method. 

\section{Conclusions} \label{sec:conclusions}
The goal of this work is to assess the efficiency of the algorithm proposed in Paper 1 under more realistic conditions, i.e., using LSST light curves with observational properties representative of the survey, a physically motivated period distribution for the underlying binary population, and a broader range of binary signals. To do so, we analysed 20000 light curves per band, half of them containing pure-noise DRW realisations and half including periodic modulations generated by different combinations of physical processes. Specifically, in this work we considered periodic light curves whose signal has been generated from Doppler boosting, periodic feeding and self-lensing in different combinations. The light curves match the properties expected after the completion of the LSST survey, with a cadence reflecting the real observing LSST strategy considering both the Wide-Fast-Deep and Deep-Drilling-Field modes (Davies et al., in prep.). All sources have a light curve in each of the 6 LSST filters (ugrizy), and we account for instrument systematics, such as band-dependent SNR and outlier observations, when generating the data.

We adopted two criteria to set the detection threshold: (1) an odds ratio larger than one, requiring the binary model to be more probable than the single-source model and assuming prior knowledge of the binary fraction among LSST AGN; and (2) a fixed false-positive fraction of $f_{\rm{FP}}=10^{-3}$. With these, we assessed the efficiency of the algorithm, introduced in section~\ref{sec:methodology}, by computing the fraction of correctly identified periodic sources in different period bins spanning a range from a few days to 10 years (i.e., the expected LSST baseline). We find that the efficiency decreases as the considered period increases. This is in agreement with what is reported in Paper 1 and previous studies, where a higher number of observed cycles implies an easier detection of a true periodic signal. The main difference from Paper 1 is the true positive fraction retrieved in this work. We recover a smaller $f_{\rm{TP}}$ as reported in Tables~\ref{tab:odds_ratio_TP_FP} and ~\ref{tab:fractions} for both thresholds. Such a smaller fraction can be explained by several factors. First, here we consider each band separately instead of assuming a single idealised light curve with high cadence and small photometric errors, resulting in fewer observations, sparser cadence and lower SNR. Second, the light curves considered in this work present a realistic SNR that varies between bands. Third, the periodic signals considered here are significantly more complex than the simple sinusoidal and sawtooth signals analysed in Paper 1.

Once the efficiency of the algorithm is found for all the bands and for the two kinds of thresholds considered in this work, we predict the distribution of detectable binaries under different assumptions for the underlying period distribution. The resulting distribution is set by the competition between the increasing number of binaries towards longer periods and the decreasing efficiency of the algorithm. We identify two behaviours depending on how fast the efficiency decreases. If the efficiency decreases faster than the increase in the number of sources in each period bin, a clear peak is identified in the detected fraction distribution as a function of the period. Specifically, such a peak is identified at periods ranging between 5 and 10 years, depending on the considered band and threshold. This implies that the most likely binaries to be found by this specific detection algorithm, considering the full survey of LSST and allowing the search to such periods, will show fewer than two complete cycles in the observed light curve. If the efficiency instead decreases more slowly than the rise in the number of binaries due to the underlying population, the detected binary fraction will follow the distribution, identifying a higher fraction of binaries at higher periods. For bands with low SNR or few pointings, the efficiency of the algorithm is dominated by the assumed prior, meaning that the data do not provide enough information to dominate the posterior. 

Finally, we studied how the results change assuming alternative period distributions for the underlying population, i.e. a population of binaries not necessarily on circular orbits and whose evolution is not dominated by GW alone. This is due to the high uncertainty in our current understanding of the processes dominating the evolution of the binary separation at different stages of the dynamical evolution leading the binary to merger. To do so, we considered period distributions going as $P^{1/3}$ and $P^{8/3}$ rather than $P^{5/3}$ expected for circular binaries where the evolution is dominated by GW emission. We find that the shape of the detected binary fraction strongly depends on the shape of the underlying population. Specifically, considering the $r$ band, the position of the peak moves to higher and higher values as the exponent of the period distribution increases. These results indicate that detecting a population of MBHBs in LSST can provide a probe of their still uncertain evolution.

We plan to improve the analysis performed in this work by taking into consideration the correlations between both the noise and periodic signal hyperparameters across different bands and different instruments through a multi-output GP analysis. Such an approach allows for the coherent analysis of light curves acquired with different instruments, opening the possibility of confirming the periodic candidate with most probably one-to-a-few periods in LSST by increasing the time coverage substantially using earlier photometric surveys, such as the Zwicky Transient Facility \citep[ZTF, see][]{Bellm2019} and Gaia \citep[see][]{Gaia2016}
\begin{acknowledgements}
LB acknowledges ISCRA for awarding this project access to the LEONARDO supercomputer, owned by the EuroHPC Joint Undertaking, hosted by CINECA (Italy). LB wishes to thank the "Summer School for Astrostatistics in Crete" for providing training on the statistical methods adopted in this work. BS acknowledges the support of the Italian National PhD Programme in Space Science and Technology, based at the University of Trento (TN, Italy) - Cycle XLI. PS acknowledge the financial support from the ASI agreements n. 2024-36-HH.1-2025 and 2025-29-HH.0 and from INAF Bando di Ricerca fondamentale 2024: "The Quest for dual and binary massive black holes in the gravitational wave era" 1.05.24.01.02.
MD and RB acknowledge support from the ICSC National Research Center funded by NextGenerationEU, and financial support by the Italian Space Agency grant Phase B2/C activity for LISA mission, Agreement n.2024-NAZ-0102/PE. MC is funded by the European Union (ERC-StG-2023, MMMonsters, 101117624).
ED was supported by funding from the WSU Office of Research through a New Faculty Seed Grant.
FR acknowledges support from ELSA Euclid Legacy Science Advanced analysis tools (Grant Agreement no. 101135203) funded by the European Union.

\end{acknowledgements}
\bibliographystyle{aa} 
\bibliography{bibliography}

@article{Vehtari2021,
  author  = {Aki Vehtari and Andrew Gelman and Daniel Simpson and Bob Carpenter and Paul-Christian B{\"u}rkner},
  title   = {Bayesian Workflow},
  journal = {Bayesian Analysis},
  volume  = {16},
  number  = {2},
  pages   = {667--718},
  year    = {2021},
  doi     = {10.1214/20-BA1221},
  url     = {https://doi.org/10.1214/20-BA1221}
}

@ARTICLE{Izquierdo,
       author = {{Izquierdo-Villalba}, David and {Habouzit}, Melanie and {Bonetti}, Matteo and {Bonoli}, Silvia and {Gualandris}, Alessia and {Volonteri}, Marta and {Angeloni}, Federico and {Barausse}, Enrico and {Bhowmick}, Aklant and {Blecha}, Laura and {Bonilla Rivera}, Alexander and {Bortolas}, Elisa and {Caliskan}, Mesut and {Capelo}, Pedro R. and {Caramete}, Ana and {Caramete}, Laurentiu and {Chen}, Nianyi and {Colpi}, Monica and {Contini}, Thierry and {Dav{\'e}}, Romeel and {Dayal}, Pratika and {DeGraf}, Colin and {Deane}, Roger and {Decarli}, Roberto and {Delpech}, R{\'e}mi and {Di Matteo}, Tiziana and {Dong-P{\'a}ez}, Chi An and {Graham}, Alister W. and {Haggard}, Daryl and {Irodotou}, Dimitrios and {Johansson}, Peter H. and {Keitaanranta}, Atte and {Kelley}, Luke Zoltan and {Khan}, Fazeel Mahmood and {Langen}, Vivienne and {Li}, Kunyang and {Liao}, Shihong and {Mangiagli}, Alberto and {Marsat}, Sylvain and {McCaffrey}, Joe and {Ni}, Yueying and {Pillay}, Coral and {Pislan}, Florentina-Crenguta and {Rawlings}, Alex and {Regan}, John and {Reinoso}, Basti{\'a}n and {Roth}, Jaelyn and {Ruiz}, Milton and {Sergijenko}, Olga and {Sesana}, Alberto and {Shaifullah}, Golam and {Singh}, Jasbir and {Spinoso}, Daniele and {Toubiana}, Alexandre and {Tremmel}, Michael and {Trinca}, Alessandro and {Valiante}, Rosa and {Zhou}, Yihao and {Dubois}, Yohan and {Graziani}, Luca and {Lovell}, Christopher C. and {Peirani}, Sebastien and {Roper}, William J. and {Schaye}, Joop and {Schneider}, Raffaella and {Trebitsch}, Maxime and {Vijayan}, Aswin and {Vogelsberger}, Mark and {Wilkins}, Stephen and {Wise}, John},
        title = "{The LISA Astrophysics MBHcatalogues Project: A comparison of predictions of simulated massive black hole binaries}",
      journal = {arXiv e-prints},
         year = 2026,
        month = apr,
          eid = {arXiv:2605.00092},
        pages = {arXiv:2605.00092},
          doi = {10.48550/arXiv.2605.00092},
archivePrefix = {arXiv},
       eprint = {2605.00092},
 primaryClass = {astro-ph.GA},
       adsurl = {https://ui.adsabs.harvard.edu/abs/2026arXiv260500092I}
}

@ARTICLE{Begelman1980,
       author = {{Begelman}, M.~C. and {Blandford}, R.~D. and {Rees}, M.~J.},
        title = "{Massive black hole binaries in active galactic nuclei}",
      journal = {\nat},
         year = 1980,
        month = sep,
       volume = {287},
       number = {5780},
        pages = {307-309},
          doi = {10.1038/287307a0},
       adsurl = {https://ui.adsabs.harvard.edu/abs/1980Natur.287..307B}
}

@ARTICLE{Dotti2012,
       author = {{Dotti}, M. and {Sesana}, A. and {Decarli}, R.},
        title = "{Massive Black Hole Binaries: Dynamical Evolution and Observational Signatures}",
      journal = {Advances in Astronomy},
         year = 2012,
        month = jan,
       volume = {2012},
          eid = {940568},
        pages = {940568},
          doi = {10.1155/2012/940568},
archivePrefix = {arXiv},
       eprint = {1111.0664},
 primaryClass = {astro-ph.CO},
       adsurl = {https://ui.adsabs.harvard.edu/abs/2012AdAst2012E...3D}
}

@ARTICLE{lisa1,
       author = {{Amaro-Seoane}, Pau and {Audley}, Heather and {Babak}, Stanislav and {Baker}, John and {Barausse}, Enrico and {Bender}, Peter and {Berti}, Emanuele and {Binetruy}, Pierre and {Born}, Michael and {Bortoluzzi}, Daniele and {Camp}, Jordan and {Caprini}, Chiara and {Cardoso}, Vitor and {Colpi}, Monica and {Conklin}, John and {Cornish}, Neil and {Cutler}, Curt and {Danzmann}, Karsten and {Dolesi}, Rita and {Ferraioli}, Luigi and {Ferroni}, Valerio and {Fitzsimons}, Ewan and {Gair}, Jonathan and {Gesa Bote}, Lluis and {Giardini}, Domenico and {Gibert}, Ferran and {Grimani}, Catia and {Halloin}, Hubert and {Heinzel}, Gerhard and {Hertog}, Thomas and {Hewitson}, Martin and {Holley-Bockelmann}, Kelly and {Hollington}, Daniel and {Hueller}, Mauro and {Inchauspe}, Henri and {Jetzer}, Philippe and {Karnesis}, Nikos and {Killow}, Christian and {Klein}, Antoine and {Klipstein}, Bill and {Korsakova}, Natalia and {Larson}, Shane L and {Livas}, Jeffrey and {Lloro}, Ivan and {Man}, Nary and {Mance}, Davor and {Martino}, Joseph and {Mateos}, Ignacio and {McKenzie}, Kirk and {McWilliams}, Sean T and {Miller}, Cole and {Mueller}, Guido and {Nardini}, Germano and {Nelemans}, Gijs and {Nofrarias}, Miquel and {Petiteau}, Antoine and {Pivato}, Paolo and {Plagnol}, Eric and {Porter}, Ed and {Reiche}, Jens and {Robertson}, David and {Robertson}, Norna and {Rossi}, Elena and {Russano}, Giuliana and {Schutz}, Bernard and {Sesana}, Alberto and {Shoemaker}, David and {Slutsky}, Jacob and {Sopuerta}, Carlos F. and {Sumner}, Tim and {Tamanini}, Nicola and {Thorpe}, Ira and {Troebs}, Michael and {Vallisneri}, Michele and {Vecchio}, Alberto and {Vetrugno}, Daniele and {Vitale}, Stefano and {Volonteri}, Marta and {Wanner}, Gudrun and {Ward}, Harry and {Wass}, Peter and {Weber}, William and {Ziemer}, John and {Zweifel}, Peter},
        title = "{Laser Interferometer Space Antenna}",
      journal = {arXiv e-prints},
         year = 2017,
        month = feb,
          eid = {arXiv:1702.00786},
        pages = {arXiv:1702.00786},
archivePrefix = {arXiv},
       eprint = {1702.00786},
 primaryClass = {astro-ph.IM},
       adsurl = {https://ui.adsabs.harvard.edu/abs/2017arXiv170200786A}
}

@ARTICLE{lisa2,
       author = {{Amaro-Seoane}, Pau and {Andrews}, Jeff and {Arca Sedda}, Manuel and {Askar}, Abbas and {Baghi}, Quentin and {Balasov}, Razvan and {Bartos}, Imre and {Bavera}, Simone S. and {Bellovary}, Jillian and {Berry}, Christopher P.~L. and {Berti}, Emanuele and {Bianchi}, Stefano and {Blecha}, Laura and {Blondin}, St{\'e}phane and {Bogdanovi{\'c}}, Tamara and {Boissier}, Samuel and {Bonetti}, Matteo and {Bonoli}, Silvia and {Bortolas}, Elisa and {Breivik}, Katelyn and {Capelo}, Pedro R. and {Caramete}, Laurentiu and {Cattorini}, Federico and {Charisi}, Maria and {Chaty}, Sylvain and {Chen}, Xian and {Chru{\'s}li{\'n}ska}, Martyna and {Chua}, Alvin J.~K. and {Church}, Ross and {Colpi}, Monica and {D'Orazio}, Daniel and {Danielski}, Camilla and {Davies}, Melvyn B. and {Dayal}, Pratika and {De Rosa}, Alessandra and {Derdzinski}, Andrea and {Destounis}, Kyriakos and {Dotti}, Massimo and {Dutan}, Ioana and {Dvorkin}, Irina and {Fabj}, Gaia and {Foglizzo}, Thierry and {Ford}, Saavik and {Fouvry}, Jean-Baptiste and {Franchini}, Alessia and {Fragos}, Tassos and {Fryer}, Chris and {Gaspari}, Massimo and {Gerosa}, Davide and {Graziani}, Luca and {Groot}, Paul and {Habouzit}, Melanie and {Haggard}, Daryl and {Haiman}, Zoltan and {Han}, Wen-Biao and {Istrate}, Alina and {Johansson}, Peter H. and {Khan}, Fazeel Mahmood and {Kimpson}, Tomas and {Kokkotas}, Kostas and {Kong}, Albert and {Korol}, Valeriya and {Kremer}, Kyle and {Kupfer}, Thomas and {Lamberts}, Astrid and {Larson}, Shane and {Lau}, Mike and {Liu}, Dongliang and {Lloyd-Ronning}, Nicole and {Lodato}, Giuseppe and {Lupi}, Alessandro and {Ma}, Chung-Pei and {Maccarone}, Tomas and {Mandel}, Ilya and {Mangiagli}, Alberto and {Mapelli}, Michela and {Mathis}, St{\'e}phane and {Mayer}, Lucio and {McGee}, Sean and {McKernan}, Barry and {Miller}, M. Coleman and {Mota}, David F. and {Mumpower}, Matthew and {Nasim}, Syeda S. and {Nelemans}, Gijs and {Noble}, Scott and {Pacucci}, Fabio and {Panessa}, Francesca and {Paschalidis}, Vasileios and {Pfister}, Hugo and {Porquet}, Delphine and {Quenby}, John and {Ricarte}, Angelo and {R{\"o}pke}, Friedrich K. and {Regan}, John and {Rosswog}, Stephan and {Ruiter}, Ashley and {Ruiz}, Milton and {Runnoe}, Jessie and {Schneider}, Raffaella and {Schnittman}, Jeremy and {Secunda}, Amy and {Sesana}, Alberto and {Seto}, Naoki and {Shao}, Lijing and {Shapiro}, Stuart and {Sopuerta}, Carlos and {Stone}, Nicholas C. and {Suvorov}, Arthur and {Tamanini}, Nicola and {Tamfal}, Tomas and {Tauris}, Thomas and {Temmink}, Karel and {Tomsick}, John and {Toonen}, Silvia and {Torres-Orjuela}, Alejandro and {Toscani}, Martina and {Tsokaros}, Antonios and {Unal}, Caner and {V{\'a}zquez-Aceves}, Ver{\'o}nica and {Valiante}, Rosa and {van Putten}, Maurice and {van Roestel}, Jan and {Vignali}, Christian and {Volonteri}, Marta and {Wu}, Kinwah and {Younsi}, Ziri and {Yu}, Shenghua and {Zane}, Silvia and {Zwick}, Lorenz and {Antonini}, Fabio and {Baibhav}, Vishal and {Barausse}, Enrico and {Bonilla Rivera}, Alexander and {Branchesi}, Marica and {Branduardi-Raymont}, Graziella and {Burdge}, Kevin and {Chakraborty}, Srija and {Cuadra}, Jorge and {Dage}, Kristen and {Davis}, Benjamin and {de Mink}, Selma E. and {Decarli}, Roberto and {Doneva}, Daniela and {Escoffier}, Stephanie and {Gandhi}, Poshak and {Haardt}, Francesco and {Lousto}, Carlos O. and {Nissanke}, Samaya and {Nordhaus}, Jason and {O'Shaughnessy}, Richard and {Portegies Zwart}, Simon and {Pound}, Adam and {Schussler}, Fabian and {Sergijenko}, Olga and {Spallicci}, Alessandro and {Vernieri}, Daniele and {Vigna-G{\'o}mez}, Alejandro},
        title = "{Astrophysics with the Laser Interferometer Space Antenna}",
      journal = {Living Reviews in Relativity},
         year = 2023,
        month = dec,
       volume = {26},
       number = {1},
          eid = {2},
        pages = {2},
          doi = {10.1007/s41114-022-00041-y},
archivePrefix = {arXiv},
       eprint = {2203.06016},
 primaryClass = {gr-qc},
       adsurl = {https://ui.adsabs.harvard.edu/abs/2023LRR....26....2A}
}

@ARTICLE{pta,
       author = {{Verbiest}, J.~P.~W. and {Lentati}, L. and {Hobbs}, G. and {van Haasteren}, R. and {Demorest}, P.~B. and {Janssen}, G.~H. and {Wang}, J. -B. and {Desvignes}, G. and {Caballero}, R.~N. and {Keith}, M.~J. and {Champion}, D.~J. and {Arzoumanian}, Z. and {Babak}, S. and {Bassa}, C.~G. and {Bhat}, N.~D.~R. and {Brazier}, A. and {Brem}, P. and {Burgay}, M. and {Burke-Spolaor}, S. and {Chamberlin}, S.~J. and {Chatterjee}, S. and {Christy}, B. and {Cognard}, I. and {Cordes}, J.~M. and {Dai}, S. and {Dolch}, T. and {Ellis}, J.~A. and {Ferdman}, R.~D. and {Fonseca}, E. and {Gair}, J.~R. and {Garver-Daniels}, N.~E. and {Gentile}, P. and {Gonzalez}, M.~E. and {Graikou}, E. and {Guillemot}, L. and {Hessels}, J.~W.~T. and {Jones}, G. and {Karuppusamy}, R. and {Kerr}, M. and {Kramer}, M. and {Lam}, M.~T. and {Lasky}, P.~D. and {Lassus}, A. and {Lazarus}, P. and {Lazio}, T.~J.~W. and {Lee}, K.~J. and {Levin}, L. and {Liu}, K. and {Lynch}, R.~S. and {Lyne}, A.~G. and {Mckee}, J. and {McLaughlin}, M.~A. and {McWilliams}, S.~T. and {Madison}, D.~R. and {Manchester}, R.~N. and {Mingarelli}, C.~M.~F. and {Nice}, D.~J. and {Os{\l}owski}, S. and {Palliyaguru}, N.~T. and {Pennucci}, T.~T. and {Perera}, B.~B.~P. and {Perrodin}, D. and {Possenti}, A. and {Petiteau}, A. and {Ransom}, S.~M. and {Reardon}, D. and {Rosado}, P.~A. and {Sanidas}, S.~A. and {Sesana}, A. and {Shaifullah}, G. and {Shannon}, R.~M. and {Siemens}, X. and {Simon}, J. and {Smits}, R. and {Spiewak}, R. and {Stairs}, I.~H. and {Stappers}, B.~W. and {Stinebring}, D.~R. and {Stovall}, K. and {Swiggum}, J.~K. and {Taylor}, S.~R. and {Theureau}, G. and {Tiburzi}, C. and {Toomey}, L. and {Vallisneri}, M. and {van Straten}, W. and {Vecchio}, A. and {Wang}, Y. and {Wen}, L. and {You}, X.~P. and {Zhu}, W.~W. and {Zhu}, X. -J.},
        title = "{The International Pulsar Timing Array: First data release}",
      journal = {\mnras},
         year = 2016,
        month = may,
       volume = {458},
       number = {2},
        pages = {1267-1288},
          doi = {10.1093/mnras/stw347},
archivePrefix = {arXiv},
       eprint = {1602.03640},
 primaryClass = {astro-ph.IM},
       adsurl = {https://ui.adsabs.harvard.edu/abs/2016MNRAS.458.1267V}
}

@ARTICLE{Antoniadis23,
       author = {{EPTA Collaboration} and {InPTA Collaboration} and {Antoniadis}, J. and {Arumugam}, P. and {Arumugam}, S. and {Babak}, S. and {Bagchi}, M. and {Bak Nielsen}, A. -S. and {Bassa}, C.~G. and {Bathula}, A. and {Berthereau}, A. and {Bonetti}, M. and {Bortolas}, E. and {Brook}, P.~R. and {Burgay}, M. and {Caballero}, R.~N. and {Chalumeau}, A. and {Champion}, D.~J. and {Chanlaridis}, S. and {Chen}, S. and {Cognard}, I. and {Dandapat}, S. and {Deb}, D. and {Desai}, S. and {Desvignes}, G. and {Dhanda-Batra}, N. and {Dwivedi}, C. and {Falxa}, M. and {Ferdman}, R.~D. and {Franchini}, A. and {Gair}, J.~R. and {Goncharov}, B. and {Gopakumar}, A. and {Graikou}, E. and {Grie{\ss}meier}, J. -M. and {Guillemot}, L. and {Guo}, Y.~J. and {Gupta}, Y. and {Hisano}, S. and {Hu}, H. and {Iraci}, F. and {Izquierdo-Villalba}, D. and {Jang}, J. and {Jawor}, J. and {Janssen}, G.~H. and {Jessner}, A. and {Joshi}, B.~C. and {Kareem}, F. and {Karuppusamy}, R. and {Keane}, E.~F. and {Keith}, M.~J. and {Kharbanda}, D. and {Kikunaga}, T. and {Kolhe}, N. and {Kramer}, M. and {Krishnakumar}, M.~A. and {Lackeos}, K. and {Lee}, K.~J. and {Liu}, K. and {Liu}, Y. and {Lyne}, A.~G. and {McKee}, J.~W. and {Maan}, Y. and {Main}, R.~A. and {Mickaliger}, M.~B. and {Ni{\c{t}}u}, I.~C. and {Nobleson}, K. and {Paladi}, A.~K. and {Parthasarathy}, A. and {Perera}, B.~B.~P. and {Perrodin}, D. and {Petiteau}, A. and {Porayko}, N.~K. and {Possenti}, A. and {Prabu}, T. and {Quelquejay Leclere}, H. and {Rana}, P. and {Samajdar}, A. and {Sanidas}, S.~A. and {Sesana}, A. and {Shaifullah}, G. and {Singha}, J. and {Speri}, L. and {Spiewak}, R. and {Srivastava}, A. and {Stappers}, B.~W. and {Surnis}, M. and {Susarla}, S.~C. and {Susobhanan}, A. and {Takahashi}, K. and {Tarafdar}, P. and {Theureau}, G. and {Tiburzi}, C. and {van der Wateren}, E. and {Vecchio}, A. and {Venkatraman Krishnan}, V. and {Verbiest}, J.~P.~W. and {Wang}, J. and {Wang}, L. and {Wu}, Z.},
        title = "{The second data release from the European Pulsar Timing Array. III. Search for gravitational wave signals}",
      journal = {\aap},
         year = 2023,
        month = oct,
       volume = {678},
          eid = {A50},
        pages = {A50},
          doi = {10.1051/0004-6361/202346844},
archivePrefix = {arXiv},
       eprint = {2306.16214},
 primaryClass = {astro-ph.HE},
       adsurl = {https://ui.adsabs.harvard.edu/abs/2023A\&A...678A..50E}
}

@article{Agazie23,
doi = {10.3847/2041-8213/acdac6},
url = {https://dx.doi.org/10.3847/2041-8213/acdac6},
year = {2023},
month = {jun},
publisher = {The American Astronomical Society},
volume = {951},
number = {1},
pages = {L8},
author = {Agazie, Gabriella and Anumarlapudi, Akash and Archibald, Anne M. and Arzoumanian, Zaven and Baker, Paul T. and Bécsy, Bence and Blecha, Laura and Brazier, Adam and Brook, Paul R. and Burke-Spolaor, Sarah and Burnette, Rand and Case, Robin and Charisi, Maria and Chatterjee, Shami and Chatziioannou, Katerina and Cheeseboro, Belinda D. and Chen, Siyuan and Cohen, Tyler and Cordes, James M. and Cornish, Neil J. and Crawford, Fronefield and Cromartie, H. Thankful and Crowter, Kathryn and Cutler, Curt J. and DeCesar, Megan E. and DeGan, Dallas and Demorest, Paul B. and Deng, Heling and Dolch, Timothy and Drachler, Brendan and Ellis, Justin A. and Ferrara, Elizabeth C. and Fiore, William and Fonseca, Emmanuel and Freedman, Gabriel E. and Garver-Daniels, Nate and Gentile, Peter A. and Gersbach, Kyle A. and Glaser, Joseph and Good, Deborah C. and Gültekin, Kayhan and Hazboun, Jeffrey S. and Hourihane, Sophie and Islo, Kristina and Jennings, Ross J. and Johnson, Aaron D. and Jones, Megan L. and Kaiser, Andrew R. and Kaplan, David L. and Kelley, Luke Zoltan and Kerr, Matthew and Key, Joey S. and Klein, Tonia C. and Laal, Nima and Lam, Michael T. and Lamb, William G. and W. Lazio, T. Joseph and Lewandowska, Natalia and Littenberg, Tyson B. and Liu, Tingting and Lommen, Andrea and Lorimer, Duncan R. and Luo, Jing and Lynch, Ryan S. and Ma, Chung-Pei and Madison, Dustin R. and Mattson, Margaret A. and McEwen, Alexander and McKee, James W. and McLaughlin, Maura A. and McMann, Natasha and Meyers, Bradley W. and Meyers, Patrick M. and Mingarelli, Chiara M. F. and Mitridate, Andrea and Natarajan, Priyamvada and Ng, Cherry and Nice, David J. and Ocker, Stella Koch and Olum, Ken D. and Pennucci, Timothy T. and Perera, Benetge B. P. and Petrov, Polina and Pol, Nihan S. and Radovan, Henri A. and Ransom, Scott M. and Ray, Paul S. and Romano, Joseph D. and Sardesai, Shashwat C. and Schmiedekamp, Ann and Schmiedekamp, Carl and Schmitz, Kai and Schult, Levi and Shapiro-Albert, Brent J. and Siemens, Xavier and Simon, Joseph and Siwek, Magdalena S. and Stairs, Ingrid H. and Stinebring, Daniel R. and Stovall, Kevin and Sun, Jerry P. and Susobhanan, Abhimanyu and Swiggum, Joseph K. and Taylor, Jacob and Taylor, Stephen R. and Turner, Jacob E. and Unal, Caner and Vallisneri, Michele and van Haasteren, Rutger and Vigeland, Sarah J. and Wahl, Haley M. and Wang, Qiaohong and Witt, Caitlin A. and Young, Olivia and The NANOGrav Collaboration},
title = {The NANOGrav 15 yr Data Set: Evidence for a Gravitational-wave Background},
journal = {The Astrophysical Journal Letters}
}

@article{Reardon23,
   title={Search for an Isotropic Gravitational-wave Background with the Parkes Pulsar Timing Array},
   volume={951},
   ISSN={2041-8213},
   url={http://dx.doi.org/10.3847/2041-8213/acdd02},
   DOI={10.3847/2041-8213/acdd02},
   number={1},
   journal={The Astrophysical Journal Letters},
   publisher={American Astronomical Society},
   author={Reardon, Daniel J. and Zic, Andrew and Shannon, Ryan M. and Hobbs, George B. and Bailes, Matthew and Di Marco, Valentina and Kapur, Agastya and Rogers, Axl F. and Thrane, Eric and Askew, Jacob and Bhat, N. D. Ramesh and Cameron, Andrew and Curyło, Małgorzata and Coles, William A. and Dai, Shi and Goncharov, Boris and Kerr, Matthew and Kulkarni, Atharva and Levin, Yuri and Lower, Marcus E. and Manchester, Richard N. and Mandow, Rami and Miles, Matthew T. and Nathan, Rowina S. and Osłowski, Stefan and Russell, Christopher J. and Spiewak, Renée and Zhang, Songbo and Zhu, Xing-Jiang},
   year={2023},
   month=jun, pages={L6} }

@article{Xu23,
   title={Searching for the Nano-Hertz Stochastic Gravitational Wave Background with the Chinese Pulsar Timing Array Data Release I},
   volume={23},
   ISSN={1674-4527},
   url={http://dx.doi.org/10.1088/1674-4527/acdfa5},
   DOI={10.1088/1674-4527/acdfa5},
   number={7},
   journal={Research in Astronomy and Astrophysics},
   publisher={IOP Publishing},
   author={Xu, Heng and Chen, Siyuan and Guo, Yanjun and Jiang, Jinchen and Wang, Bojun and Xu, Jiangwei and Xue, Zihan and Nicolas Caballero, R. and Yuan, Jianping and Xu, Yonghua and Wang, Jingbo and Hao, Longfei and Luo, Jingtao and Lee, Kejia and Han, Jinlin and Jiang, Peng and Shen, Zhiqiang and Wang, Min and Wang, Na and Xu, Renxin and Wu, Xiangping and Manchester, Richard and Qian, Lei and Guan, Xin and Huang, Menglin and Sun, Chun and Zhu, Yan},
   year={2023},
   month=jun, pages={075024} }

@ARTICLE{Duffel2020,
       author = {{Duffell}, Paul C. and {D'Orazio}, Daniel and {Derdzinski}, Andrea and {Haiman}, Zoltan and {MacFadyen}, Andrew and {Rosen}, Anna L. and {Zrake}, Jonathan},
        title = "{Circumbinary Disks: Accretion and Torque as a Function of Mass Ratio and Disk Viscosity}",
      journal = {\apj},
         year = 2020,
        month = sep,
       volume = {901},
       number = {1},
          eid = {25},
        pages = {25},
          doi = {10.3847/1538-4357/abab95},
archivePrefix = {arXiv},
       eprint = {1911.05506},
 primaryClass = {astro-ph.SR},
       adsurl = {https://ui.adsabs.harvard.edu/abs/2020ApJ...901...25D}
}

@ARTICLE{Derosa2019,
       author = {{De Rosa}, Alessandra and {Vignali}, Cristian and {Bogdanovi{\'c}}, Tamara and {Capelo}, Pedro R. and {Charisi}, Maria and {Dotti}, Massimo and {Husemann}, Bernd and {Lusso}, Elisabeta and {Mayer}, Lucio and {Paragi}, Zsolt and {Runnoe}, Jessie and {Sesana}, Alberto and {Steinborn}, Lisa and {Bianchi}, Stefano and {Colpi}, Monica and {del Valle}, Luciano and {Frey}, S{\'a}ndor and {Gab{\'a}nyi}, Krisztina {\'E}. and {Giustini}, Margherita and {Guainazzi}, Matteo and {Haiman}, Zoltan and {Herrera Ruiz}, Noelia and {Herrero-Illana}, Rub{\'e}n and {Iwasawa}, Kazushi and {Komossa}, S. and {Lena}, Davide and {Loiseau}, Nora and {Perez-Torres}, Miguel and {Piconcelli}, Enrico and {Volonteri}, Marta},
        title = "{The quest for dual and binary supermassive black holes: A multi-messenger view}",
      journal = {\nar},
         year = 2019,
        month = dec,
       volume = {86},
          eid = {101525},
        pages = {101525},
          doi = {10.1016/j.newar.2020.101525},
archivePrefix = {arXiv},
       eprint = {2001.06293},
 primaryClass = {astro-ph.GA},
       adsurl = {https://ui.adsabs.harvard.edu/abs/2019NewAR..8601525D}
}

@ARTICLE{DOrazio2023,
       author = {{D'Orazio}, Daniel J. and {Charisi}, Maria},
        title = "{Observational Signatures of Supermassive Black Hole Binaries}",
      journal = {arXiv e-prints},
         year = 2023,
        month = oct,
          eid = {arXiv:2310.16896},
        pages = {arXiv:2310.16896},
          doi = {10.48550/arXiv.2310.16896},
archivePrefix = {arXiv},
       eprint = {2310.16896},
 primaryClass = {astro-ph.HE},
       adsurl = {https://ui.adsabs.harvard.edu/abs/2023arXiv231016896D}
}

@ARTICLE{Bogdanovic2021,
       author = {{Bogdanovic}, Tamara and {Miller}, M. Coleman and {Blecha}, Laura},
        title = "{Electromagnetic Counterparts to Massive Black Hole Mergers}",
      journal = {arXiv e-prints},
         year = 2021,
        month = sep,
          eid = {arXiv:2109.03262},
        pages = {arXiv:2109.03262},
archivePrefix = {arXiv},
       eprint = {2109.03262},
 primaryClass = {astro-ph.HE},
       adsurl = {https://ui.adsabs.harvard.edu/abs/2021arXiv210903262B}
}

@article{Ivezic2019,
   title={LSST: From Science Drivers to Reference Design and Anticipated Data Products},
   volume={873},
   ISSN={1538-4357},
   url={http://dx.doi.org/10.3847/1538-4357/ab042c},
   DOI={10.3847/1538-4357/ab042c},
   number={2},
   journal={The Astrophysical Journal},
   publisher={American Astronomical Society},
   author={{Ivezi{\'c}}, {\v{Z}}eljko and Kahn, Steven M. and Tyson, J. Anthony and Abel, Bob and Acosta, Emily and Allsman, Robyn and Alonso, David and AlSayyad, Yusra and Anderson, Scott F. and Andrew, John and P. Angel, James Roger and Angeli, George Z. and Ansari, Reza and Antilogus, Pierre and Araujo, Constanza and Armstrong, Robert and Arndt, Kirk T. and Astier, Pierre and Aubourg, Éric and Auza, Nicole and Axelrod, Tim S. and Bard, Deborah J. and Barr, Jeff D. and Barrau, Aurelian and Bartlett, James G. and Bauer, Amanda E. and Bauman, Brian J. and Baumont, Sylvain and Bechtol, Ellen and Bechtol, Keith and Becker, Andrew C. and Becla, Jacek and Beldica, Cristina and Bellavia, Steve and Bianco, Federica B. and Biswas, Rahul and Blanc, Guillaume and Blazek, Jonathan and Blandford, Roger D. and Bloom, Josh S. and Bogart, Joanne and Bond, Tim W. and Booth, Michael T. and Borgland, Anders W. and Borne, Kirk and Bosch, James F. and Boutigny, Dominique and Brackett, Craig A. and Bradshaw, Andrew and Brandt, William Nielsen and Brown, Michael E. and Bullock, James S. and Burchat, Patricia and Burke, David L. and Cagnoli, Gianpietro and Calabrese, Daniel and Callahan, Shawn and Callen, Alice L. and Carlin, Jeffrey L. and Carlson, Erin L. and Chandrasekharan, Srinivasan and Charles-Emerson, Glenaver and Chesley, Steve and Cheu, Elliott C. and Chiang, Hsin-Fang and Chiang, James and Chirino, Carol and Chow, Derek and Ciardi, David R. and Claver, Charles F. and Cohen-Tanugi, Johann and Cockrum, Joseph J. and Coles, Rebecca and Connolly, Andrew J. and Cook, Kem H. and Cooray, Asantha and Covey, Kevin R. and Cribbs, Chris and Cui, Wei and Cutri, Roc and Daly, Philip N. and Daniel, Scott F. and Daruich, Felipe and Daubard, Guillaume and Daues, Greg and Dawson, William and Delgado, Francisco and Dellapenna, Alfred and Peyster, Robert de and Val-Borro, Miguel de and Digel, Seth W. and Doherty, Peter and Dubois, Richard and Dubois-Felsmann, Gregory P. and Durech, Josef and Economou, Frossie and Eifler, Tim and Eracleous, Michael and Emmons, Benjamin L. and Neto, Angelo Fausti and Ferguson, Henry and Figueroa, Enrique and Fisher-Levine, Merlin and Focke, Warren and Foss, Michael D. and Frank, James and Freemon, Michael D. and Gangler, Emmanuel and Gawiser, Eric and Geary, John C. and Gee, Perry and Geha, Marla and Gessner, Charles J. B. and Gibson, Robert R. and Gilmore, D. Kirk and Glanzman, Thomas and Glick, William and Goldina, Tatiana and Goldstein, Daniel A. and Goodenow, Iain and Graham, Melissa L. and Gressler, William J. and Gris, Philippe and Guy, Leanne P. and Guyonnet, Augustin and Haller, Gunther and Harris, Ron and Hascall, Patrick A. and Haupt, Justine and Hernandez, Fabio and Herrmann, Sven and Hileman, Edward and Hoblitt, Joshua and Hodgson, John A. and Hogan, Craig and Howard, James D. and Huang, Dajun and Huffer, Michael E. and Ingraham, Patrick and Innes, Walter R. and Jacoby, Suzanne H. and Jain, Bhuvnesh and Jammes, Fabrice and Jee, M. James and Jenness, Tim and Jernigan, Garrett and Jevremović, Darko and Johns, Kenneth and Johnson, Anthony S. and Johnson, Margaret W. G. and Jones, R. Lynne and Juramy-Gilles, Claire and Jurić, Mario and Kalirai, Jason S. and Kallivayalil, Nitya J. and Kalmbach, Bryce and Kantor, Jeffrey P. and Karst, Pierre and Kasliwal, Mansi M. and Kelly, Heather and Kessler, Richard and Kinnison, Veronica and Kirkby, David and Knox, Lloyd and Kotov, Ivan V. and Krabbendam, Victor L. and Krughoff, K. Simon and Kubánek, Petr and Kuczewski, John and Kulkarni, Shri and Ku, John and Kurita, Nadine R. and Lage, Craig S. and Lambert, Ron and Lange, Travis and Langton, J. Brian and Guillou, Laurent Le and Levine, Deborah and Liang, Ming and Lim, Kian-Tat and Lintott, Chris J. and Long, Kevin E. and Lopez, Margaux and Lotz, Paul J. and Lupton, Robert H. and Lust, Nate B. and MacArthur, Lauren A. and Mahabal, Ashish and Mandelbaum, Rachel and Markiewicz, Thomas W. and Marsh, Darren S. and Marshall, Philip J. and Marshall, Stuart and May, Morgan and McKercher, Robert and McQueen, Michelle and Meyers, Joshua and Migliore, Myriam and Miller, Michelle and Mills, David J. and Miraval, Connor and Moeyens, Joachim and Moolekamp, Fred E. and Monet, David G. and Moniez, Marc and Monkewitz, Serge and Montgomery, Christopher and Morrison, Christopher B. and Mueller, Fritz and Muller, Gary P. and Arancibia, Freddy Muñoz and Neill, Douglas R. and Newbry, Scott P. and Nief, Jean-Yves and Nomerotski, Andrei and Nordby, Martin and O’Connor, Paul and Oliver, John and Olivier, Scot S. and Olsen, Knut and O’Mullane, William and Ortiz, Sandra and Osier, Shawn and Owen, Russell E. and Pain, Reynald and Palecek, Paul E. and Parejko, John K. and Parsons, James B. and Pease, Nathan M. and Peterson, J. Matt and Peterson, John R. and Petravick, Donald L. and Petrick, M. E. Libby and Petry, Cathy E. and Pierfederici, Francesco and Pietrowicz, Stephen and Pike, Rob and Pinto, Philip A. and Plante, Raymond and Plate, Stephen and Plutchak, Joel P. and Price, Paul A. and Prouza, Michael and Radeka, Veljko and Rajagopal, Jayadev and Rasmussen, Andrew P. and Regnault, Nicolas and Reil, Kevin A. and Reiss, David J. and Reuter, Michael A. and Ridgway, Stephen T. and Riot, Vincent J. and Ritz, Steve and Robinson, Sean and Roby, William and Roodman, Aaron and Rosing, Wayne and Roucelle, Cecille and Rumore, Matthew R. and Russo, Stefano and Saha, Abhijit and Sassolas, Benoit and Schalk, Terry L. and Schellart, Pim and Schindler, Rafe H. and Schmidt, Samuel and Schneider, Donald P. and Schneider, Michael D. and Schoening, William and Schumacher, German and Schwamb, Megan E. and Sebag, Jacques and Selvy, Brian and Sembroski, Glenn H. and Seppala, Lynn G. and Serio, Andrew and Serrano, Eduardo and Shaw, Richard A. and Shipsey, Ian and Sick, Jonathan and Silvestri, Nicole and Slater, Colin T. and Smith, J. Allyn and Smith, R. Chris and Sobhani, Shahram and Soldahl, Christine and Storrie-Lombardi, Lisa and Stover, Edward and Strauss, Michael A. and Street, Rachel A. and Stubbs, Christopher W. and Sullivan, Ian S. and Sweeney, Donald and Swinbank, John D. and Szalay, Alexander and Takacs, Peter and Tether, Stephen A. and Thaler, Jon J. and Thayer, John Gregg and Thomas, Sandrine and Thornton, Adam J. and Thukral, Vaikunth and Tice, Jeffrey and Trilling, David E. and Turri, Max and Berg, Richard Van and Berk, Daniel Vanden and Vetter, Kurt and Virieux, Francoise and Vucina, Tomislav and Wahl, William and Walkowicz, Lucianne and Walsh, Brian and Walter, Christopher W. and Wang, Daniel L. and Wang, Shin-Yawn and Warner, Michael and Wiecha, Oliver and Willman, Beth and Winters, Scott E. and Wittman, David and Wolff, Sidney C. and Wood-Vasey, W. Michael and Wu, Xiuqin and Xin, Bo and Yoachim, Peter and Zhan, Hu},
   year={2019},
   month=Mar, pages={111} }

@ARTICLE{haiman23,
       author = {{Haiman}, Zolt{\'a}n and {Xin}, Chengcheng and {Bogdanovi{\'c}}, Tamara and {Amaro Seoane}, Pau and {Bonetti}, Matteo and {Casey-Clyde}, J. Andrew and {Charisi}, Maria and {Colpi}, Monica and {Davelaar}, Jordy and {De Rosa}, Alessandra and {D'Orazio}, Daniel J. and {Futrowsky}, Kate and {Gandhi}, Poshak and {Graham}, Alister W. and {Greene}, Jenny E. and {Habouzit}, Melanie and {Haggard}, Daryl and {Holley-Bockelmann}, Kelly and {Liu}, Xin and {Mangiagli}, Alberto and {Mastrobuono-Battisti}, Alessandra and {McGee}, Sean and {Mingarelli}, Chiara M.~F. and {Nemmen}, Rodrigo and {Palmese}, Antonella and {Porquet}, Delphine and {Sesana}, Alberto and {Stemo}, Aaron and {Torres-Orjuela}, Alejandro and {Zrake}, Jonathan},
        title = "{Massive Black Hole Binaries as LISA Precursors in the Roman High Latitude Time Domain Survey}",
      journal = {arXiv e-prints},
         year = 2023,
        month = jun,
          eid = {arXiv:2306.14990},
        pages = {arXiv:2306.14990},
          doi = {10.48550/arXiv.2306.14990},
archivePrefix = {arXiv},
       eprint = {2306.14990},
 primaryClass = {astro-ph.HE},
       adsurl = {https://ui.adsabs.harvard.edu/abs/2023arXiv230614990H}
}

@ARTICLE{MacFayden2008,
   author = {{MacFadyen}, A.~I. and {Milosavljevi{\'c}}, M.},
    title = "{An Eccentric Circumbinary Accretion Disk and the Detection of Binary Massive Black Holes}",
  journal = {\apj},
   eprint = {astro-ph/0607467},
     year = 2008,
    month = jan,
   volume = 672,
    pages = {83-93},
      doi = {10.1086/523869},
   adsurl = {http://adsabs.harvard.edu/abs/2008ApJ...672...83M}
}

@ARTICLE{Hayasaki2008,
   author = {{Hayasaki}, K. and {Mineshige}, S. and {Ho}, L.~C.},
    title = "{A Supermassive Binary Black Hole with Triple Disks}",
  journal = {\apj},
archivePrefix = "arXiv",
   eprint = {0708.2555},
     year = 2008,
    month = aug,
   volume = 682,
    pages = {1134-1140},
      doi = {10.1086/588837},
   adsurl = {http://adsabs.harvard.edu/abs/2008ApJ...682.1134H}
}

@ARTICLE{Kelley21,
       author = {{Kelley}, Luke Zoltan and {D'Orazio}, Daniel J. and {Di Stefano}, Rosanne},
        title = "{Gravitational self-lensing in populations of massive black hole binaries}",
      journal = {\mnras},
         year = 2021,
        month = dec,
       volume = {508},
       number = {2},
        pages = {2524-2536},
          doi = {10.1093/mnras/stab2776},
archivePrefix = {arXiv},
       eprint = {2107.07522},
 primaryClass = {astro-ph.HE},
       adsurl = {https://ui.adsabs.harvard.edu/abs/2021MNRAS.508.2524K}
}

@ARTICLE{DOrazio2018,
       author = {{D'Orazio}, Daniel J. and {Di Stefano}, Rosanne},
        title = "{Periodic self-lensing from accreting massive black hole binaries}",
      journal = {\mnras},
         year = 2018,
        month = mar,
       volume = {474},
       number = {3},
        pages = {2975-2986},
          doi = {10.1093/mnras/stx2936},
archivePrefix = {arXiv},
       eprint = {1707.02335},
 primaryClass = {astro-ph.HE},
       adsurl = {https://ui.adsabs.harvard.edu/abs/2018MNRAS.474.2975D}
}

@ARTICLE{DOrazio2015,
   author = {{D'Orazio}, D.~J. and {Haiman}, Z. and {Schiminovich}, D.},
    title = "{Relativistic boost as the cause of periodicity in a massive black-hole binary candidate}",
  journal = {\nat},
archivePrefix = "arXiv",
   eprint = {1509.04301},
 primaryClass = "astro-ph.HE",
     year = 2015,
    month = sep,
   volume = 525,
    pages = {351-353},
      doi = {10.1038/nature15262},
   adsurl = {http://adsabs.harvard.edu/abs/2015Natur.525..351D}
}

@ARTICLE{Charisi2018,
       author = {{Charisi}, Maria and {Haiman}, Zolt{\'a}n and {Schiminovich}, David and {D'Orazio}, Daniel J.},
        title = "{Testing the relativistic Doppler boost hypothesis for supermassive black hole binary candidates}",
      journal = {\mnras},
         year = 2018,
        month = jun,
       volume = {476},
       number = {4},
        pages = {4617-4628},
          doi = {10.1093/mnras/sty516},
archivePrefix = {arXiv},
       eprint = {1801.06189},
 primaryClass = {astro-ph.GA},
       adsurl = {https://ui.adsabs.harvard.edu/abs/2018MNRAS.476.4617C}
}

@article{Ulrich1997,
   author = "Ulrich, Marie-Helene and Maraschi, Laura and Urry, C. Megan",
   title = "VARIABILITY OF ACTIVE GALACTIC NUCLEI", 
   journal= "Annual Review of Astronomy and Astrophysics",
   year = "1997",
   volume = "35",
   number = "Volume 35, 1997",
   pages = "445-502",
   doi = "https://doi.org/10.1146/annurev.astro.35.1.445",
   url = "https://www.annualreviews.org/content/journals/10.1146/annurev.astro.35.1.445",
   publisher = "Annual Reviews",
   issn = "1545-4282",
   type = "Journal Article",
  }

@article{Vaughan16, 
author = { S. Vaughan and P. Uttley and A. G. Markowitz and D. Huppenkothen and M. J. Middleton and W. N. Alston and J. D. Scargle and W. M. Farr }, 
title = { False periodicities in quasar time-domain surveys }, 
eprint = { Arxiv:1606.02620v1 }, 
journal = { Monthly Notices of the Royal Astronomical Society },
volume = { 461 },
pages = { 3145 },
year = { 2016 },
url = { https://arxiv.org/abs/1606.02620 }, 
}

@article{Lin2026,
doi = {10.3847/1538-4357/ae29a7},
url = {https://doi.org/10.3847/1538-4357/ae29a7},
year = {2026},
month = {jan},
publisher = {The American Astronomical Society},
volume = {997},
number = {2},
pages = {316},
author = {Lin, Allison and Charisi, Maria and Haiman, Zoltán},
title = {Lomb–Scargle Periodogram Struggles with Non-sinusoidal Supermassive Black Hole Binary Signatures in Quasar Lightcurves},
journal = {The Astrophysical Journal}
}

@article{Hu2020,
   title={Spikey: self-lensing flares from eccentric SMBH binaries},
   volume={495},
   ISSN={1365-2966},
   url={http://dx.doi.org/10.1093/mnras/staa1312},
   DOI={10.1093/mnras/staa1312},
   number={4},
   journal={Monthly Notices of the Royal Astronomical Society},
   publisher={Oxford University Press (OUP)},
   author={Hu, Betty X and D’Orazio, Daniel J and Haiman, Zoltán and Smith, Krista Lynne and Snios, Bradford and Charisi, Maria and Di Stefano, Rosanne},
   year={2020},
   month=may, pages={4061–4070} }

@article{Zrake_2021,
   title={Equilibrium Eccentricity of Accreting Binaries},
   volume={909},
   ISSN={2041-8213},
   url={http://dx.doi.org/10.3847/2041-8213/abdd1c},
   DOI={10.3847/2041-8213/abdd1c},
   number={1},
   journal={The Astrophysical Journal Letters},
   publisher={American Astronomical Society},
   author={Zrake, Jonathan and Tiede, Christopher and MacFadyen, Andrew and Haiman, Zoltán},
   year={2021},
   month=mar, pages={L13} }

@article{Westernacher_Schneider22,
   title={Multiband light curves from eccentric accreting supermassive black hole binaries},
   volume={106},
   ISSN={2470-0029},
   url={http://dx.doi.org/10.1103/PhysRevD.106.103010},
   DOI={10.1103/physrevd.106.103010},
   number={10},
   journal={Physical Review D},
   publisher={American Physical Society (APS)},
   author={Westernacher-Schneider, John Ryan and Zrake, Jonathan and MacFadyen, Andrew and Haiman, Zoltán},
   year={2022},
   month=nov }

@article{Cocchiararo24,
   title={Electromagnetic signatures from accreting massive black hole binaries in time domain photometric surveys},
   volume={691},
   ISSN={1432-0746},
   url={http://dx.doi.org/10.1051/0004-6361/202449598},
   DOI={10.1051/0004-6361/202449598},
   journal={Astronomy \&amp; Astrophysics},
   publisher={EDP Sciences},
   author={Cocchiararo, F. and Franchini, A. and Lupi, A. and Sesana, A.},
   year={2024},
   month=nov, pages={A250} }

@misc{bertassi2026,
      title={Identification of periodicities with arbitrary shapes in AGN light curves}, 
      author={Lorenzo Bertassi and Maria Charisi and Riccardo Buscicchio and Fabio Rigamonti and Jessie Runnoe and Massimo Dotti},
      year={2026},
      eprint={2512.13688},
      archivePrefix={arXiv},
      primaryClass={astro-ph.GA},
      url={https://arxiv.org/abs/2512.13688}, 
}

@BOOK{Rasmussen2006,
       author = {{Rasmussen}, Carl Edward and {Williams}, Christopher K.~I.},
        title = "{Gaussian Processes for Machine Learning}",
         year = 2006,
       adsurl = {https://ui.adsabs.harvard.edu/abs/2006gpml.book.....R}
}

@article{Skilling2006,
author = {John Skilling},
title = {{Nested sampling for general Bayesian computation}},
volume = {1},
journal = {Bayesian Analysis},
number = {4},
publisher = {International Society for Bayesian Analysis},
pages = {833 -- 859},
year = {2006},
doi = {10.1214/06-BA127},
URL = {https://doi.org/10.1214/06-BA127}
}

@ARTICLE{VanderPlas18,
       author = {{VanderPlas}, Jacob T.},
        title = "{Understanding the Lomb-Scargle Periodogram}",
      journal = {\apjs},
         year = 2018,
        month = may,
       volume = {236},
       number = {1},
          eid = {16},
        pages = {16},
          doi = {10.3847/1538-4365/aab766},
archivePrefix = {arXiv},
       eprint = {1703.09824},
 primaryClass = {astro-ph.IM},
       adsurl = {https://ui.adsabs.harvard.edu/abs/2018ApJS..236...16V}
}

@ARTICLE{MacLeod10,
       author = {{MacLeod}, C.~L. and {Ivezi{\'c}}, {\v{Z}}. and {Kochanek}, C.~S. and {Koz{\l}owski}, S. and {Kelly}, B. and {Bullock}, E. and {Kimball}, A. and {Sesar}, B. and {Westman}, D. and {Brooks}, K. and {Gibson}, R. and {Becker}, A.~C. and {de Vries}, W.~H.},
        title = "{Modeling the Time Variability of SDSS Stripe 82 Quasars as a Damped Random Walk}",
      journal = {\apj},
         year = 2010,
        month = oct,
       volume = {721},
       number = {2},
        pages = {1014-1033},
          doi = {10.1088/0004-637X/721/2/1014},
archivePrefix = {arXiv},
       eprint = {1004.0276},
 primaryClass = {astro-ph.CO},
       adsurl = {https://ui.adsabs.harvard.edu/abs/2010ApJ...721.1014M}
}

@ARTICLE{Kelly2009,
       author = {{Kelly}, Brandon C. and {Bechtold}, Jill and {Siemiginowska}, Aneta},
        title = "{Are the Variations in Quasar Optical Flux Driven by Thermal Fluctuations?}",
      journal = {\apj},
         year = 2009,
        month = jun,
       volume = {698},
       number = {1},
        pages = {895-910},
          doi = {10.1088/0004-637X/698/1/895},
archivePrefix = {arXiv},
       eprint = {0903.5315},
 primaryClass = {astro-ph.CO},
       adsurl = {https://ui.adsabs.harvard.edu/abs/2009ApJ...698..895K}
}

@misc{albert2020,
      title={JAXNS: a high-performance nested sampling package based on JAX}, 
      author={Joshua G. Albert},
      year={2020},
      eprint={2012.15286},
      archivePrefix={arXiv},
      primaryClass={astro-ph.IM},
      url={https://arxiv.org/abs/2012.15286}, 
}

@software{tinygp,
  author       = {Foreman-Mackey, Daniel and
                  Weixiang Yu and
                  Yadav, Sachin and
                  Becker, McCoy Reynolds and
                  Caplar, Neven and
                  Huppenkothen, Daniela and
                  Killestein, Thomas and
                  Tronsgaard, René and
                  Rashid, Theo and
                  Schmerler, Steve},
  title        = {dfm/tinygp: The tiniest of Gaussian Process
                   libraries
                  },
  month        = mar,
  year         = 2026,
  publisher    = {Zenodo},
  version      = {v0.3.1rc1},
  doi          = {10.5281/zenodo.19035146},
  url          = {https://doi.org/10.5281/zenodo.19035146},
  swhid        = {swh:1:dir:6ca1b08190b451b55c4f5b1d6440bc67953fe3ac
                   ;origin=https://doi.org/10.5281/zenodo.6389737;vis
                   it=swh:1:snp:a1f5a1a5223649ea5fbc61ecfd19c26566aa2
                   8a8;anchor=swh:1:rel:129d94a54bf7c649a2e82e444056c
                   cf0b68357ac;path=dfm-tinygp-c7637bb
                  },
}

@misc{dorazio2024,
      title={Fast Methods for Computing Photometric Variability of Eccentric Binaries: Boosting, Lensing, and Variable Accretion}, 
      author={Daniel J. D'Orazio and Paul C. Duffell and Christopher Tiede},
      year={2024},
      eprint={2403.05629},
      archivePrefix={arXiv},
      primaryClass={astro-ph.HE},
      url={https://arxiv.org/abs/2403.05629}, 
}

@article{Robnik24,
    author = {Robnik, Jakob and Bayer, Adrian E and Charisi, Maria and Haiman, Zoltán and Lin, Allison and Seljak, Uroš},
    title = {Periodicity significance testing with null-signal templates: reassessment of PTF’s SMBH binary candidates},
    journal = {Monthly Notices of the Royal Astronomical Society},
    volume = {534},
    number = {2},
    pages = {1609-1620},
    year = {2024},
    month = {10},
    issn = {0035-8711},
    doi = {10.1093/mnras/stae2220},
    url = {https://doi.org/10.1093/mnras/stae2220},
    eprint = {https://academic.oup.com/mnras/article-pdf/534/2/1609/59577103/stae2220.pdf},
}

@ARTICLE{Covino2022,
       author = {{Covino}, Stefano and {Tobar}, Felipe and {Treves}, Aldo},
        title = "{Detecting the periodicity of highly irregularly sampled light curves with Gaussian processes: the case of SDSS J025214.67-002813.7}",
      journal = {\mnras},
         year = 2022,
        month = jun,
       volume = {513},
       number = {2},
        pages = {2841-2849},
          doi = {10.1093/mnras/stac596},
archivePrefix = {arXiv},
       eprint = {2203.03614},
 primaryClass = {astro-ph.IM},
       adsurl = {https://ui.adsabs.harvard.edu/abs/2022MNRAS.513.2841C}
}

@ARTICLE{rigamonti25,
       author = {{Rigamonti}, Fabio and {Bertassi}, Lorenzo and {Buscicchio}, Riccardo and {Cocchiararo}, Fabiola and {Covino}, Stefano and {Dotti}, Massimo and {Sesana}, Alberto and {Severgnini}, Paola},
        title = "{Variability in the massive black hole binary candidate SDSS J2320+0024: no evidence for periodic modulation}",
      journal = {arXiv e-prints},
         year = 2025,
        month = may,
          eid = {arXiv:2505.22706},
        pages = {arXiv:2505.22706},
          doi = {10.48550/arXiv.2505.22706},
archivePrefix = {arXiv},
       eprint = {2505.22706},
 primaryClass = {astro-ph.GA},
       adsurl = {https://ui.adsabs.harvard.edu/abs/2025arXiv250522706R}
}

@ARTICLE{Park2026,
       author = {{Park}, Kevin and {Haiman}, Zolt{\'a}n and {Xin}, Chengcheng and {Shen}, Tzuken and {Villar}, Ashley and {Davelaar}, Jordy},
        title = "{Self-lensing flares from black hole binaries. V. Systematic searches in LSST}",
      journal = {\prd},
         year = 2026,
        month = feb,
       volume = {113},
       number = {4},
          eid = {043055},
        pages = {043055},
          doi = {10.1103/p328-62sl},
archivePrefix = {arXiv},
       eprint = {2512.08427},
 primaryClass = {astro-ph.HE},
       adsurl = {https://ui.adsabs.harvard.edu/abs/2026PhRvD.113d3055P}
}

@ARTICLE{Foustoul2025,
       author = {{Foustoul}, V. and {Webb}, N.~A. and {Mignon-Risse}, R. and {Kammoun}, E. and {Volonteri}, M. and {Dong-P{\'a}ez}, C.~A.},
        title = "{A catalogue of candidate milliparsec-separation massive black hole binaries from long-term optical photometric monitoring}",
      journal = {\aap},
         year = 2025,
        month = jul,
       volume = {699},
          eid = {A55},
        pages = {A55},
          doi = {10.1051/0004-6361/202452818},
archivePrefix = {arXiv},
       eprint = {2505.06656},
 primaryClass = {astro-ph.GA},
       adsurl = {https://ui.adsabs.harvard.edu/abs/2025A&A...699A..55F}
}

@software{Bradbury2018,
  author = {James Bradbury and Roy Frostig and Peter Hawkins and Matthew James Johnson and Yash Katariya and Chris Leary and Dougal Maclaurin and George Necula and Adam Paszke and Jake Vander{P}las and Skye Wanderman-{M}ilne and Qiao Zhang},
  title = {{JAX}: composable transformations of {P}ython+{N}um{P}y programs},
  url = {http://github.com/jax-ml/jax},
  version = {0.3.13},
  year = {2018},
}

@article{Waddell2024,
   title={The eROSITA Final Equatorial Depth Survey (eFEDS): Complex absorption and soft excesses in hard X-ray–selected active galactic nuclei},
   volume={690},
   ISSN={1432-0746},
   url={http://dx.doi.org/10.1051/0004-6361/202245572},
   DOI={10.1051/0004-6361/202245572},
   journal={Astronomy \& amp; Astrophysics},
   publisher={EDP Sciences},
   author={Waddell, S. G. H. and Nandra, K. and Buchner, J. and Wu, Q. and Shen, Y. and Arcodia, R. and Merloni, A. and Salvato, M. and Dauser, T. and Boller, Th. and Liu, T. and Comparat, J. and Wolf, J. and Dwelly, T. and Ricci, C. and Brownstein, J. R. and Brusa, M.},
   year={2024},
   month=Oct, pages={A132} }

@ARTICLE{Bellm2019,
       author = {{Bellm}, Eric C. and {Kulkarni}, Shrinivas R. and {Graham}, Matthew J. and {Dekany}, Richard and {Smith}, Roger M. and {Riddle}, Reed and {Masci}, Frank J. and {Helou}, George and {Prince}, Thomas A. and {Adams}, Scott M. and {Barbarino}, C. and {Barlow}, Tom and {Bauer}, James and {Beck}, Ron and {Belicki}, Justin and {Biswas}, Rahul and {Blagorodnova}, Nadejda and {Bodewits}, Dennis and {Bolin}, Bryce and {Brinnel}, Valery and {Brooke}, Tim and {Bue}, Brian and {Bulla}, Mattia and {Burruss}, Rick and {Cenko}, S. Bradley and {Chang}, Chan-Kao and {Connolly}, Andrew and {Coughlin}, Michael and {Cromer}, John and {Cunningham}, Virginia and {De}, Kishalay and {Delacroix}, Alex and {Desai}, Vandana and {Duev}, Dmitry A. and {Eadie}, Gwendolyn and {Farnham}, Tony L. and {Feeney}, Michael and {Feindt}, Ulrich and {Flynn}, David and {Franckowiak}, Anna and {Frederick}, S. and {Fremling}, C. and {Gal-Yam}, Avishay and {Gezari}, Suvi and {Giomi}, Matteo and {Goldstein}, Daniel A. and {Golkhou}, V. Zach and {Goobar}, Ariel and {Groom}, Steven and {Hacopians}, Eugean and {Hale}, David and {Henning}, John and {Ho}, Anna Y.~Q. and {Hover}, David and {Howell}, Justin and {Hung}, Tiara and {Huppenkothen}, Daniela and {Imel}, David and {Ip}, Wing-Huen and {Ivezi{\'c}}, {\v{Z}}eljko and {Jackson}, Edward and {Jones}, Lynne and {Juric}, Mario and {Kasliwal}, Mansi M. and {Kaspi}, S. and {Kaye}, Stephen and {Kelley}, Michael S.~P. and {Kowalski}, Marek and {Kramer}, Emily and {Kupfer}, Thomas and {Landry}, Walter and {Laher}, Russ R. and {Lee}, Chien-De and {Lin}, Hsing Wen and {Lin}, Zhong-Yi and {Lunnan}, Ragnhild and {Giomi}, Matteo and {Mahabal}, Ashish and {Mao}, Peter and {Miller}, Adam A. and {Monkewitz}, Serge and {Murphy}, Patrick and {Ngeow}, Chow-Choong and {Nordin}, Jakob and {Nugent}, Peter and {Ofek}, Eran and {Patterson}, Maria T. and {Penprase}, Bryan and {Porter}, Michael and {Rauch}, Ludwig and {Rebbapragada}, Umaa and {Reiley}, Dan and {Rigault}, Mickael and {Rodriguez}, Hector and {van Roestel}, Jan and {Rusholme}, Ben and {van Santen}, Jakob and {Schulze}, S. and {Shupe}, David L. and {Singer}, Leo P. and {Soumagnac}, Maayane T. and {Stein}, Robert and {Surace}, Jason and {Sollerman}, Jesper and {Szkody}, Paula and {Taddia}, F. and {Terek}, Scott and {Van Sistine}, Angela and {van Velzen}, Sjoert and {Vestrand}, W. Thomas and {Walters}, Richard and {Ward}, Charlotte and {Ye}, Quan-Zhi and {Yu}, Po-Chieh and {Yan}, Lin and {Zolkower}, Jeffry},
        title = "{The Zwicky Transient Facility: System Overview, Performance, and First Results}",
      journal = {\pasp},
         year = 2019,
        month = jan,
       volume = {131},
       number = {995},
        pages = {018002},
          doi = {10.1088/1538-3873/aaecbe},
archivePrefix = {arXiv},
       eprint = {1902.01932},
 primaryClass = {astro-ph.IM},
       adsurl = {https://ui.adsabs.harvard.edu/abs/2019PASP..131a8002B}
}

@ARTICLE{Gaia2016,
       author = {{Gaia Collaboration} and {Prusti}, T. and {de Bruijne}, J.~H.~J. and {Brown}, A.~G.~A. and {Vallenari}, A. and {Babusiaux}, C. and {Bailer-Jones}, C.~A.~L. and {Bastian}, U. and {Biermann}, M. and {Evans}, D.~W. and {Eyer}, L. and {Jansen}, F. and {Jordi}, C. and {Klioner}, S.~A. and {Lammers}, U. and {Lindegren}, L. and {Luri}, X. and {Mignard}, F. and {Milligan}, D.~J. and {Panem}, C. and {Poinsignon}, V. and {Pourbaix}, D. and {Randich}, S. and {Sarri}, G. and {Sartoretti}, P. and {Siddiqui}, H.~I. and {Soubiran}, C. and {Valette}, V. and {van Leeuwen}, F. and {Walton}, N.~A. and {Aerts}, C. and {Arenou}, F. and {Cropper}, M. and {Drimmel}, R. and {H{\o}g}, E. and {Katz}, D. and {Lattanzi}, M.~G. and {O'Mullane}, W. and {Grebel}, E.~K. and {Holland}, A.~D. and {Huc}, C. and {Passot}, X. and {Bramante}, L. and {Cacciari}, C. and {Casta{\~n}eda}, J. and {Chaoul}, L. and {Cheek}, N. and {De Angeli}, F. and {Fabricius}, C. and {Guerra}, R. and {Hern{\'a}ndez}, J. and {Jean-Antoine-Piccolo}, A. and {Masana}, E. and {Messineo}, R. and {Mowlavi}, N. and {Nienartowicz}, K. and {Ord{\'o}{\~n}ez-Blanco}, D. and {Panuzzo}, P. and {Portell}, J. and {Richards}, P.~J. and {Riello}, M. and {Seabroke}, G.~M. and {Tanga}, P. and {Th{\'e}venin}, F. and {Torra}, J. and {Els}, S.~G. and {Gracia-Abril}, G. and {Comoretto}, G. and {Garcia-Reinaldos}, M. and {Lock}, T. and {Mercier}, E. and {Altmann}, M. and {Andrae}, R. and {Astraatmadja}, T.~L. and {Bellas-Velidis}, I. and {Benson}, K. and {Berthier}, J. and {Blomme}, R. and {Busso}, G. and {Carry}, B. and {Cellino}, A. and {Clementini}, G. and {Cowell}, S. and {Creevey}, O. and {Cuypers}, J. and {Davidson}, M. and {De Ridder}, J. and {de Torres}, A. and {Delchambre}, L. and {Dell'Oro}, A. and {Ducourant}, C. and {Fr{\'e}mat}, Y. and {Garc{\'\i}a-Torres}, M. and {Gosset}, E. and {Halbwachs}, J.-L. and {Hambly}, N.~C. and {Harrison}, D.~L. and {Hauser}, M. and {Hestroffer}, D. and {Hodgkin}, S.~T. and {Huckle}, H.~E. and {Hutton}, A. and {Jasniewicz}, G. and {Jordan}, S. and {Kontizas}, M. and {Korn}, A.~J. and {Lanzafame}, A.~C. and {Manteiga}, M. and {Moitinho}, A. and {Muinonen}, K. and {Osinde}, J. and {Pancino}, E. and {Pauwels}, T. and {Petit}, J.-M. and {Recio-Blanco}, A. and {Robin}, A.~C. and {Sarro}, L.~M. and {Siopis}, C. and {Smith}, M. and {Smith}, K.~W. and {Sozzetti}, A. and {Thuillot}, W. and {van Reeven}, W. and {Viala}, Y. and {Abbas}, U. and {Abreu Aramburu}, A. and {Accart}, S. and {Aguado}, J.~J. and {Allan}, P.~M. and {Allasia}, W. and {Altavilla}, G. and {{\'A}lvarez}, M.~A. and {Alves}, J. and {Anderson}, R.~I. and {Andrei}, A.~H. and {Anglada Varela}, E. and {Antiche}, E. and {Antoja}, T. and {Ant{\'o}n}, S. and {Arcay}, B. and {Atzei}, A. and {Ayache}, L. and {Bach}, N. and {Baker}, S.~G. and {Balaguer-N{\'u}{\~n}ez}, L. and {Barache}, C. and {Barata}, C. and {Barbier}, A. and {Barblan}, F. and {Baroni}, M. and {Barrado y Navascu{\'e}s}, D. and {Barros}, M. and {Barstow}, M.~A. and {Becciani}, U. and {Bellazzini}, M. and {Bellei}, G. and {Bello Garc{\'\i}a}, A. and {Belokurov}, V. and {Bendjoya}, P. and {Berihuete}, A. and {Bianchi}, L. and {Bienaym{\'e}}, O. and {Billebaud}, F. and {Blagorodnova}, N. and {Blanco-Cuaresma}, S. and {Boch}, T. and {Bombrun}, A. and {Borrachero}, R. and {Bouquillon}, S. and {Bourda}, G. and {Bouy}, H. and {Bragaglia}, A. and {Breddels}, M.~A. and {Brouillet}, N. and {Br{\"u}semeister}, T. and {Bucciarelli}, B. and {Budnik}, F. and {Burgess}, P. and {Burgon}, R. and {Burlacu}, A. and {Busonero}, D. and {Buzzi}, R. and {Caffau}, E. and {Cambras}, J. and {Campbell}, H. and {Cancelliere}, R. and {Cantat-Gaudin}, T. and {Carlucci}, T. and {Carrasco}, J.~M. and {Castellani}, M. and {Charlot}, P. and {Charnas}, J. and {Charvet}, P. and {Chassat}, F. and {Chiavassa}, A. and {Clotet}, M. and {Cocozza}, G. and {Collins}, R.~S. and {Collins}, P. and {Costigan}, G.},
        title = "{The Gaia mission}",
      journal = {\aap},
         year = 2016,
        month = nov,
       volume = {595},
          eid = {A1},
        pages = {A1},
          doi = {10.1051/0004-6361/201629272},
archivePrefix = {arXiv},
       eprint = {1609.04153},
 primaryClass = {astro-ph.IM},
       adsurl = {https://ui.adsabs.harvard.edu/abs/2016A&A...595A...1G}
}

@article{Severgnini2026,
title = {A new era for dual AGN science with SHARP},
journal = {New Astronomy},
volume = {128},
pages = {102610},
year = {2026},
issn = {1384-1076},
doi = {https://doi.org/10.1016/j.newast.2026.102610},
url = {https://www.sciencedirect.com/science/article/pii/S138410762600093X},
author = {P. Severgnini and C. Vignali and A. {De Rosa} and E. Portaluri and F. Rigamonti and L. Battistini and L. Bertassi and E. Bertola and S. Bianchi and E. Bortolas and C. Cicone and Q. D’Amato and R. {Della Ceca} and I. Delvecchio and M. Dotti and J. Harms and I. Lamperti and F. Mannucci and M. Parvatikar and B. Sala and M. Scialpi and R. Serafinelli and J. Singh and M.V. Zanchettin}
}

@article{Volonteri2009,
   title={SUB-PARSEC SUPERMASSIVE BINARY QUASARS: EXPECTATIONS AT
                    <i>z</i>
                    &lt; 1},
   volume={703},
   ISSN={1538-4357},
   url={http://dx.doi.org/10.1088/0004-637X/703/1/L86},
   DOI={10.1088/0004-637x/703/1/l86},
   number={1},
   journal={The Astrophysical Journal},
   publisher={American Astronomical Society},
   author={Volonteri, M. and Miller, J. M. and Dotti, M.},
   year={2009},
   month=Sept, pages={L86–L89} }

@ARTICLE{Casey_Clyde2025,
       author = {{Casey-Clyde}, J. Andrew and {Mingarelli}, Chiara M.~F. and {Greene}, Jenny E. and {Goulding}, Andy D. and {Chen}, Siyuan and {Trump}, Jonathan R.},
        title = "{Quasars Can Signpost Supermassive Black Hole Binaries}",
      journal = {\apj},
         year = 2025,
        month = jul,
       volume = {987},
       number = {2},
          eid = {106},
        pages = {106},
          doi = {10.3847/1538-4357/adce05},
archivePrefix = {arXiv},
       eprint = {2405.19406},
 primaryClass = {astro-ph.HE},
       adsurl = {https://ui.adsabs.harvard.edu/abs/2025ApJ...987..106C}
}

@article{Paolillo2025,
   title={Continuum optical-UV and X-ray variability of AGN: current results and future challenges},
   volume={48},
   ISSN={1826-9850},
   url={http://dx.doi.org/10.1007/s40766-025-00072-5},
   DOI={10.1007/s40766-025-00072-5},
   number={8},
   journal={La Rivista del Nuovo Cimento},
   publisher={Springer Science and Business Media LLC},
   author={Paolillo, Maurizio and Papadakis, Iossif},
   year={2025},
   month=Aug, pages={537–621} }

@article{Witt2022,
   title={Quasars with Periodic Variability: Capabilities and Limitations of Bayesian Searches for Supermassive Black Hole Binaries in Time-domain Surveys},
   volume={936},
   ISSN={1538-4357},
   url={http://dx.doi.org/10.3847/1538-4357/ac8356},
   DOI={10.3847/1538-4357/ac8356},
   number={1},
   journal={The Astrophysical Journal},
   publisher={American Astronomical Society},
   author={Witt, Caitlin A. and Charisi, Maria and Taylor, Stephen R. and Burke-Spolaor, Sarah},
   year={2022},
   month=Sept, pages={89} }

@ARTICLE{Rodriguez2006,
       author = {{Rodriguez}, C. and {Taylor}, G.~B. and {Zavala}, R.~T. and {Peck}, A.~B. and {Pollack}, L.~K. and {Romani}, R.~W.},
        title = "{A Compact Supermassive Binary Black Hole System}",
      journal = {\apj},
         year = 2006,
        month = jul,
       volume = {646},
       number = {1},
        pages = {49-60},
          doi = {10.1086/504825},
archivePrefix = {arXiv},
       eprint = {astro-ph/0604042},
 primaryClass = {astro-ph},
       adsurl = {https://ui.adsabs.harvard.edu/abs/2006ApJ...646...49R}
}

@ARTICLE{Scialpi2024,
       author = {{Scialpi}, M. and {Mannucci}, F. and {Marconcini}, C. and {Venturi}, G. and {Pancino}, E. and {Marconi}, A. and {Cresci}, G. and {Belfiore}, F. and {Amiri}, A. and {Bertola}, E. and {Carniani}, S. and {Cicone}, C. and {Ciurlo}, A. and {D'Amato}, Q. and {Ginolfi}, M. and {Lusso}, E. and {Marasco}, A. and {Nardini}, E. and {Rubinur}, K. and {Severgnini}, P. and {Tozzi}, G. and {Ulivi}, L. and {Vignali}, C. and {Volonteri}, M.},
        title = "{MUSE adaptive-optics spectroscopy confirms dual active galactic nuclei and strongly lensed systems at sub-arcsec separation}",
      journal = {\aap},
         year = 2024,
        month = oct,
       volume = {690},
          eid = {A57},
        pages = {A57},
          doi = {10.1051/0004-6361/202347242},
archivePrefix = {arXiv},
       eprint = {2305.11850},
 primaryClass = {astro-ph.GA},
       adsurl = {https://ui.adsabs.harvard.edu/abs/2024A&A...690A..57S}
}

@ARTICLE{Charisi2022,
       author = {{Charisi}, Maria and {Taylor}, Stephen R. and {Runnoe}, Jessie and {Bogdanovic}, Tamara and {Trump}, Jonathan R.},
        title = "{Multimessenger time-domain signatures of supermassive black hole binaries}",
      journal = {\mnras},
         year = 2022,
        month = mar,
       volume = {510},
       number = {4},
        pages = {5929-5944},
          doi = {10.1093/mnras/stab3713},
archivePrefix = {arXiv},
       eprint = {2110.14661},
 primaryClass = {astro-ph.HE},
       adsurl = {https://ui.adsabs.harvard.edu/abs/2022MNRAS.510.5929C}
}

@ARTICLE{Banaszak2026,
       author = {{Banaszak}, Sebastian M. and {Witt}, Caitlin A. and {Miller}, Adam A.},
        title = "{Optimizing Optical Searches for Supermassive Black Hole Binaries in Active Galactic Nuclei Light Curves: Fourier versus Bayesian Periodicity Detection}",
      journal = {\apj},
         year = 2026,
        month = jun,
       volume = {1004},
       number = {1},
          eid = {25},
        pages = {25},
          doi = {10.3847/1538-4357/ae6107},
archivePrefix = {arXiv},
       eprint = {2601.15379},
 primaryClass = {astro-ph.GA},
       adsurl = {https://ui.adsabs.harvard.edu/abs/2026ApJ..1004...25B}
}

@ARTICLE{Kovacevic2026,
       author = {{Kova{\v{c}}evi{\'c}}, Andjelka B. and {Ili{\'c}}, Dragana and {Me{\v{s}}tri{\'c}}, Uro{\v{s}} and {Stankovi{\'c} Cerovi{\'c}}, Milena and {De Bonis}, Natale and {Popovi{\'c}}, Luka {\v{C}}. and {Simi{\'c}}, Sa{\v{s}}a and {Fatovi{\'c}}, Marta and {Hernandez-Garc{\'\i}a}, Lorena and {Schneider}, Donald P. and {Assef}, Roberto J. and {Panda}, Swayamtrupta and {Raiteri}, Claudia M. and {Wolf}, Christian and {Onken}, Christopher A. and {Caplar}, Neven and {Charisi}, Maria and {Ricci}, Claudio and {Xin}, Chengcheng and {Yoon}, Ilsang and {Salvato}, Mara and {Temple}, Matthew J. and {Satheesh-Sheeba}, Sarath and {De Cicco}, Demetra and {Paolillo}, Maurizio and {S{\'a}nchez-S{\'a}ez}, Paula and {Bongiorno}, Angela and {Yu}, Weixiang and {Viitanen}, Akke and {Rani}, Bindu and {Davis}, Megan C. and {Marculewicz}, Marcin and {Jurki{\'c}}, Tomislav and {Savi{\'c}}, {\DH}or{\dj}e},
        title = "{QhX Catalog of Coherent and Quasi-coherent Active Galactic Nucleus Variability from Vera C. Rubin LSST DP1}",
      journal = {\apjs},
         year = 2026,
        month = aug,
       volume = {285},
       number = {2},
          eid = {61},
        pages = {61},
          doi = {10.3847/1538-4365/ae8444},
       adsurl = {https://ui.adsabs.harvard.edu/abs/2026ApJS..285...61K}
}

@ARTICLE{Siwek_2023,
       author = {{Siwek}, Magdalena and {Weinberger}, Rainer and {Mu{\~n}oz}, Diego J. and {Hernquist}, Lars},
        title = "{Preferential accretion and circumbinary disc precession in eccentric binary systems}",
      journal = {\mnras},
         year = 2023,
        month = feb,
       volume = {518},
       number = {4},
        pages = {5059-5071},
          doi = {10.1093/mnras/stac3263},
archivePrefix = {arXiv},
       eprint = {2203.02514},
 primaryClass = {astro-ph.HE},
       adsurl = {https://ui.adsabs.harvard.edu/abs/2023MNRAS.518.5059S}
}

@MISC{eztao_code,
       author = {{Yu}, Weixiang and {Richards}, Gordon T.},
        title = "{EzTao: Easier CARMA Modeling}",
 howpublished = {Astrophysics Source Code Library, record ascl:2201.001},
         year = 2022,
        month = jan,
          eid = {ascl:2201.001},
        pages = {ascl:2201.001},
archivePrefix = {ascl},
       eprint = {2201.001},
       adsurl = {https://ui.adsabs.harvard.edu/abs/2022ascl.soft01001Y}
}

@ARTICLE{EPTA2024,
       author = {{EPTA Collaboration} and {InPTA Collaboration} and {Antoniadis}, J. and {Arumugam}, P. and {Arumugam}, S. and {Babak}, S. and {Bagchi}, M. and {Bak Nielsen}, A.-S. and {Bassa}, C.~G. and {Bathula}, A. and {Berthereau}, A. and {Bonetti}, M. and {Bortolas}, E. and {Brook}, P.~R. and {Burgay}, M. and {Caballero}, R.~N. and {Chalumeau}, A. and {Champion}, D.~J. and {Chanlaridis}, S. and {Chen}, S. and {Cognard}, I. and {Dandapat}, S. and {Deb}, D. and {Desai}, S. and {Desvignes}, G. and {Dhanda-Batra}, N. and {Dwivedi}, C. and {Falxa}, M. and {Ferdman}, R.~D. and {Franchini}, A. and {Gair}, J.~R. and {Goncharov}, B. and {Gopakumar}, A. and {Graikou}, E. and {Grie{\ss}meier}, J.-M. and {Gualandris}, A. and {Guillemot}, L. and {Guo}, Y.~J. and {Gupta}, Y. and {Hisano}, S. and {Hu}, H. and {Iraci}, F. and {Izquierdo-Villalba}, D. and {Jang}, J. and {Jawor}, J. and {Janssen}, G.~H. and {Jessner}, A. and {Joshi}, B.~C. and {Kareem}, F. and {Karuppusamy}, R. and {Keane}, E.~F. and {Keith}, M.~J. and {Kharbanda}, D. and {Kikunaga}, T. and {Kolhe}, N. and {Kramer}, M. and {Krishnakumar}, M.~A. and {Lackeos}, K. and {Lee}, K.~J. and {Liu}, K. and {Liu}, Y. and {Lyne}, A.~G. and {McKee}, J.~W. and {Maan}, Y. and {Main}, R.~A. and {Mickaliger}, M.~B. and {Ni{\c{t}}u}, I.~C. and {Nobleson}, K. and {Paladi}, A.~K. and {Parthasarathy}, A. and {Perera}, B.~B.~P. and {Perrodin}, D. and {Petiteau}, A. and {Porayko}, N.~K. and {Possenti}, A. and {Prabu}, T. and {Quelquejay Leclere}, H. and {Rana}, P. and {Samajdar}, A. and {Sanidas}, S.~A. and {Sesana}, A. and {Shaifullah}, G. and {Singha}, J. and {Speri}, L. and {Spiewak}, R. and {Srivastava}, A. and {Stappers}, B.~W. and {Surnis}, M. and {Susarla}, S.~C. and {Susobhanan}, A. and {Takahashi}, K. and {Tarafdar}, P. and {Theureau}, G. and {Tiburzi}, C. and {van der Wateren}, E. and {Vecchio}, A. and {Venkatraman Krishnan}, V. and {Verbiest}, J.~P.~W. and {Wang}, J. and {Wang}, L. and {Wu}, Z. and {Auclair}, P. and {Barausse}, E. and {Caprini}, C. and {Crisostomi}, M. and {Fastidio}, F. and {Khizriev}, T. and {Middleton}, H. and {Neronov}, A. and {Postnov}, K. and {Roper Pol}, A. and {Semikoz}, D. and {Smarra}, C. and {Steer}, D.~A. and {Truant}, R.~J. and {Valtolina}, S.},
        title = "{The second data release from the European Pulsar Timing Array. IV. Implications for massive black holes, dark matter, and the early Universe}",
      journal = {\aap},
         year = 2024,
        month = may,
       volume = {685},
          eid = {A94},
        pages = {A94},
          doi = {10.1051/0004-6361/202347433},
archivePrefix = {arXiv},
       eprint = {2306.16227},
 primaryClass = {astro-ph.CO},
       adsurl = {https://ui.adsabs.harvard.edu/abs/2024A&A...685A..94E}
}

@ARTICLE{lgwa,
       author = {{Ajith}, Parameswaran and {Seoane}, Pau Amaro and {Arca Sedda}, Manuel and {Arcodia}, Riccardo and {Badaracco}, Francesca and {Banerjee}, Biswajit and {Belgacem}, Enis and {Benetti}, Giovanni and {Benetti}, Stefano and {Bobrick}, Alexey and {Bonforte}, Alessandro and {Bortolas}, Elisa and {Braito}, Valentina and {Branchesi}, Marica and {Burrows}, Adam and {Cappellaro}, Enrico and {Della Ceca}, Roberto and {Chakraborty}, Chandrachur and {Subrahmanya}, Shreevathsa Chalathadka and {Coughlin}, Michael W. and {Covino}, Stefano and {Derdzinski}, Andrea and {Doshi}, Aayushi and {Falanga}, Maurizio and {Foffa}, Stefano and {Franchini}, Alessia and {Frigeri}, Alessandro and {Futaana}, Yoshifumi and {Gerberding}, Oliver and {Gill}, Kiranjyot and {Di Giovanni}, Matteo and {Giudice}, Ines Francesca and {Giustini}, Margherita and {Gl{\"a}ser}, Philipp and {Harms}, Jan and {van Heijningen}, Joris and {Iacovelli}, Francesco and {Kavanagh}, Bradley J. and {Kawamura}, Taichi and {Kenath}, Arun and {Keppler}, Elisabeth-Adelheid and {Kobayashi}, Chiaki and {Komatsu}, Goro and {Korol}, Valeriya and {Krishnendu}, N.~V. and {Kumar}, Prayush and {Longo}, Francesco and {Maggiore}, Michele and {Mancarella}, Michele and {Maselli}, Andrea and {Mastrobuono-Battisti}, Alessandra and {Mazzarini}, Francesco and {Melandri}, Andrea and {Melini}, Daniele and {Menina}, Sabrina and {Miniutti}, Giovanni and {Mitra}, Deeshani and {Mor{\'a}n-Fraile}, Javier and {Mukherjee}, Suvodip and {Muttoni}, Niccol{\`o} and {Olivieri}, Marco and {Onori}, Francesca and {Papa}, Maria Alessandra and {Patat}, Ferdinando and {Perali}, Andrea and {Piran}, Tsvi and {Piranomonte}, Silvia and {Pol}, Alberto Roper and {Pookkillath}, Masroor C. and {Prasad}, R. and {Prasad}, Vaishak and {De Rosa}, Alessandra and {Chowdhury}, Sourav Roy and {Serafinelli}, Roberto and {Sesana}, Alberto and {Severgnini}, Paola and {Stallone}, Angela and {Tissino}, Jacopo and {Tkal{\v{c}}i{\'c}}, Hrvoje and {Tomasella}, Lina and {Toscani}, Martina and {Vartanyan}, David and {Vignali}, Cristian and {Zaccarelli}, Lucia and {Zeoli}, Morgane and {Zuccarello}, Luciano},
        title = "{The Lunar Gravitational-wave Antenna: mission studies and science case}",
      journal = {\jcap},
         year = 2025,
        month = jan,
       volume = {2025},
       number = {1},
          eid = {108},
        pages = {108},
          doi = {10.1088/1475-7516/2025/01/108},
archivePrefix = {arXiv},
       eprint = {2404.09181},
 primaryClass = {gr-qc},
       adsurl = {https://ui.adsabs.harvard.edu/abs/2025JCAP...01..108A}
}

@ARTICLE{Singh_2026,
       author = {{Singh}, Jasbir and {Severgnini}, Paola and {Cammelli}, Vieri and {De Rosa}, Alessandra and {Vignali}, Cristian and {Rigamonti}, Fabio and {Valiante}, Rosa and {Monaco}, Pierluigi and {Tan}, Jonathan C. and {Battistini}, Lorenzo and {Della Ceca}, Roberto and {Harms}, Jan and {Parvatikar}, Manali},
        title = "{Black hole merger rates for LISA and LGWA from semi-analytical modelling of light seeds}",
      journal = {\aap},
         year = 2026,
        month = feb,
       volume = {706},
          eid = {A196},
        pages = {A196},
          doi = {10.1051/0004-6361/202556833},
archivePrefix = {arXiv},
       eprint = {2512.06094},
 primaryClass = {astro-ph.CO},
       adsurl = {https://ui.adsabs.harvard.edu/abs/2026A&A...706A.196S}
}

@article{Chen2017,
   title={Efficient computation of the gravitational wave spectrum emitted by eccentric massive black hole binaries in stellar environments},
   volume={470},
   ISSN={1365-2966},
   url={http://dx.doi.org/10.1093/mnras/stx1093},
   DOI={10.1093/mnras/stx1093},
   number={2},
   journal={Monthly Notices of the Royal Astronomical Society},
   publisher={Oxford University Press (OUP)},
   author={Chen, Siyuan and Sesana, Alberto and Del Pozzo, Walter},
   year={2017},
   month=May, pages={1738–1749} }

\begin{appendix}
\section{Evidence reweighting} 
\label{app:Z_reweight}
When dealing with models where the likelihood evaluation is computationally expensive, running nested sampling with different priors can be avoided by reweighting the evidence. This can be done if the new prior has the same support or a support contained in one of the priors used for the nested sampling run. If this is the case, the change in the evidence under prior transformation can be easily found from its definition. Assume we have the value of the evidence $Z_{\pi}$ under the initial prior $\pi(\theta)$ defined as
\begin{equation}
    Z_{\pi}=\int_{\theta} L(d|\theta) \pi(\theta) d\theta,
\end{equation}
where $d$ are the data, $\theta$ are the hyperparameters and $L(d|\theta)$ is the likelihood. Consider now a different prior $\pi'(\theta)$. The new evidence $Z_{\pi'}$ is defined as:
\begin{equation}
    Z_{\pi'}=\int_{\theta} L(d|\theta) \pi'(\theta) d\theta.
\end{equation}
Multiplying and dividing by $\pi$ and the posterior $p_{\pi}(\theta|d)$ obtained by using the initial prior one get:
\begin{equation}
    Z_{\pi'}=\int_{\theta} \frac{L(d|\theta) \ \pi(\theta)}{p_{\pi}(\theta|d)}  \frac{\pi'(\theta)}{\pi(\theta)} p_{\pi}(\theta|d) d\theta = Z_{\pi} \int_{\theta}\frac{\pi'(\theta)}{\pi(\theta)} p_{\pi}(\theta|d) d\theta,
    \label{eq:reweighted_Z}
\end{equation} 
where we used Bayes theorem to write $Z_{\pi}$ in terms of $L,\ \pi,\ p_{\pi}$. The integral on the rightmost side of Equation~\ref{eq:reweighted_Z} can be approximated by Monte Carlo integration over $p_{\pi}$. Doing so, the ratio between the evidence can be written as:
\begin{equation}
    \frac{Z_{\pi'}}{Z_{\pi}}\sim \frac{1}{N}\sum \frac{\pi'(\tilde{\theta})}{\pi(\tilde{\theta})}
\end{equation}
where $\tilde{\theta}$ are posterior samples obtained from the initial nested sampling run. 

Reweighting the evidence is a practical choice as it avoids rerunning nested sampling for each prior. We note that this approximation comes at the cost of high variance if the ratio $\pi'(\tilde{\theta})/\pi(\tilde{\theta})$ varies significantly across posterior samples, or if the number of posterior samples is insufficient. 
\end{appendix}
\end{document}